\documentclass[traditabstract]{aa} 
\usepackage{graphicx}
\usepackage{epstopdf}
\usepackage{natbib}
\usepackage{txfonts}
\usepackage{color}
\usepackage{setspace}

\usepackage{natbib,twoopt}
\usepackage[breaklinks=true]{hyperref} 
\hypersetup{
    colorlinks = true,
    citecolor = {blue}
 }

\begin{document}

\title{Estimation of a Coronal Mass Ejection Magnetic Field Strength using Radio Observations of Gyrosynchrotron Radiation}

\author{Eoin P. Carley\inst{1,2},
          Nicole Vilmer\inst{2,3},
	   Paulo J. A. Sim$\mathrm{\tilde{o}}$es\inst{4},
	   \and 
	   Br\'{i}an \'{O} Fearraigh\inst{1}
          }

\institute{Astrophysics Research Group, School of Physics, Trinity College Dublin, Dublin 2, Ireland.
			  \email{eoin.carley@tcd.ie} 
			 	\and   
LESIA, Observatoire de Paris, PSL Research University, CNRS, Sorbonne Universit\'{e}s, UPMC Univ. Paris 06, Univ. Paris Diderot, Sorbonne  Paris Cit\'{e}, 5 place Jules Janssen, 92195 Meudon, France. 
				\and
Station de Radioastronomie de Nan\c{c}ay, Observatoire de Paris, PSL Research University, CNRS, Univ. Orl\'{e}ans, 18330 Nan\c{c}ay, France. 
				\and
SUPA School of Physics and Astronomy, University of Glasgow, Glasgow G12 8QQ, United Kingdom.
                    }

 \date{Received YYYY; accepted YYYY}

 
\abstract{
Coronal mass ejections (CMEs) are large eruptions of plasma and magnetic field from the low solar corona into interplanetary space. These eruptions are often associated with the acceleration of energetic electrons which produce various sources of high intensity plasma emission. In relatively rare cases, the energetic electrons may also produce gyrosynchrotron emission from within the CME itself, allowing for a diagnostic of the CME magnetic field strength. Such a magnetic field diagnostic is important for evaluating the total magnetic energy content of the CME, which is ultimately what drives the eruption. Here we report on an unusually large source of gyrosynchrotron radiation in the form of a type IV radio burst associated with a CME occurring on 2014-September-01, observed using instrumentation from the Nan\c{c}ay Radio Astronomy Facility. A combination of spectral flux density measurements from the Nan\c{c}ay instruments and the Radio Solar Telescope Network (RSTN) from 300\,MHz to 5\,GHz reveals a gyrosynchrotron spectrum with a peak flux density at $\sim$1\,GHz. Using this radio analysis, a model for gyrosynchrotron radiation, a non-thermal electron density diagnostic using the \emph{Fermi} Gamma Ray Burst Monitor (GBM) and images of the eruption from the GOES Soft X-ray Imager (SXI), {\color{black}we are able to calculate both the magnetic field strength and the properties of the X-ray and radio emitting energetic electrons within the CME. We find the radio emission is produced by non-thermal electrons of energies $>$1\,MeV with a spectral index of $\delta$$\sim$3 in a CME magnetic field of 4.4\,G at a height of 1.3 R$_{\odot}$, while the X-ray emission is produced from a similar distribution of electrons but with much lower energies on the order of 10\,keV. We conclude by comparing the electron distribution characteristics derived from both X-ray and radio and how such an analysis can be used to define the plasma and bulk properties of a CME.}
}


\titlerunning{Estimate of Coronal Mass Ejection Magnetic Field Strength}
\authorrunning{Eoin P. Carley}   


\maketitle


\section{Introduction}
     \label{S-Introduction} 
     
Coronal mass ejections (CME) are large eruptions of plasma and magnetic field from the low solar atmosphere into the heliosphere, representing the most energetic eruptions ($>$10$^{32}$\,erg) in the solar system. Despite many years of study, the trigger and driver of such eruptions is still under investigation. Observational studies have indicated that CME magnetic energy represents the largest part of the total energy budget of the eruption \citep{emslie2004, emslie2012}. The magnetic field is also the dominant driver of the eruption early in its evolution \citep{vourlidas2000, carley2012}. However, despite having such a dominant influence on CME dynamics, little is known about CME magnetic field strength. This is due to the scarcity of measurements of the magnetic field strength of such eruptions. Therefore, any new measure of this quantity represents a rare and important diagnostic that is essential for gaining a complete picture of eruption evolution.

 \begin{figure*}[!t]
    \begin{center}
    \includegraphics[scale=0.4, trim=2cm 1cm 2cm 1cm]{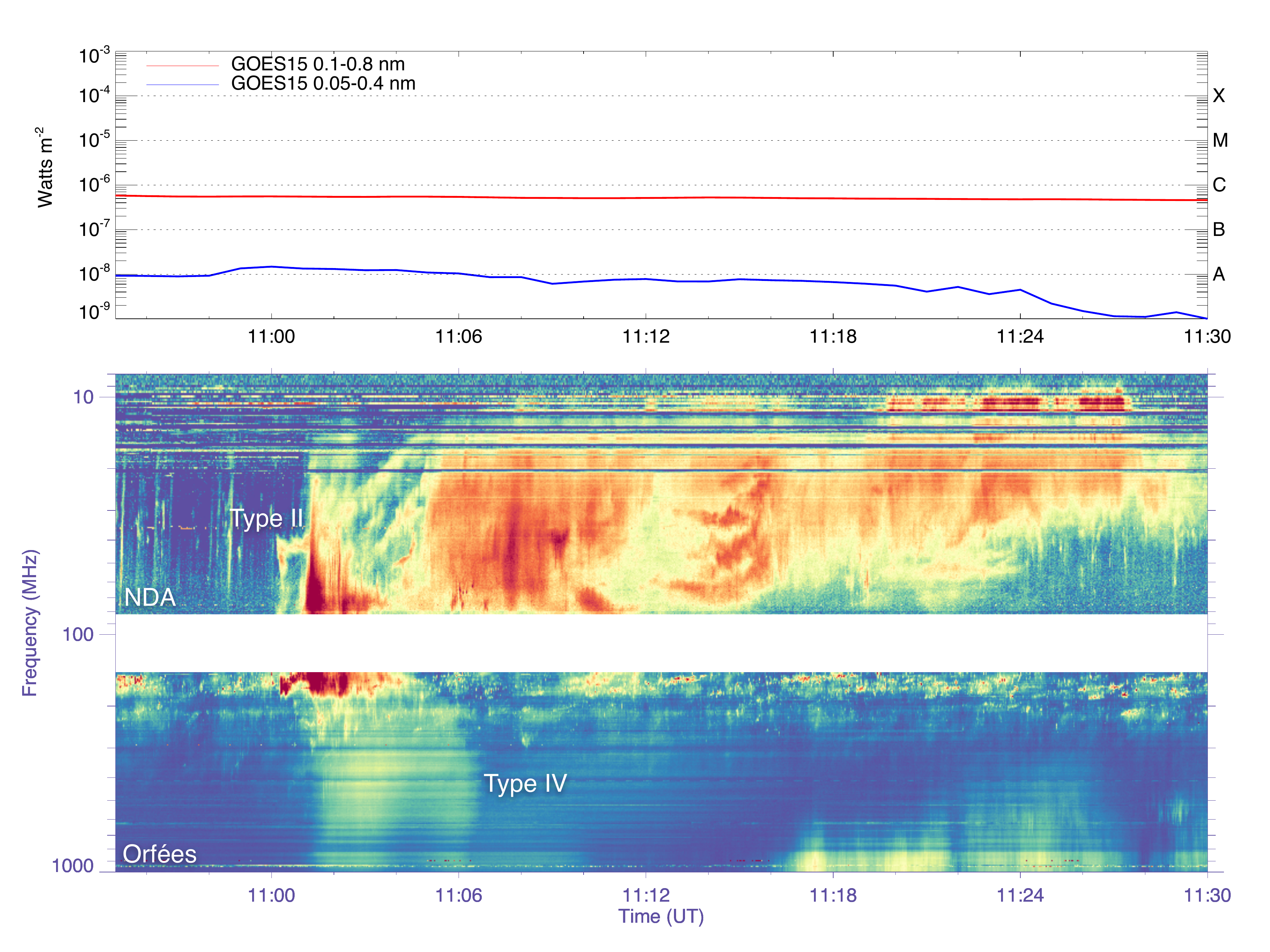}
    \caption{(a) GOES soft X-ray (SXR) time profile between 10:55--11:30\,UT, showing no significant emission as the event occurred $\sim$36$^{\circ}$ beyond the east limb. (b) NDA and Orf\'{e}es dynamic spectra from 10--1000\,MHz. A variety of radio bursts occur during the event, beginning with a type II radio burst observed in NDA (as indicated), followed by a series of complex emissions. In Orf\'{e}es a weak, broadband and smoothly varying emission is observed from 200--1000\,MHz between $\sim$11:01--11:06\,UT, which we label here as a type IV burst. The analysis in this paper concentrates on this time range and radio burst.}
    \label{fig:goes_dam_orfees}
    \end{center}
\end{figure*}    

Magnetic field strength measurements of coronal ejecta have historically been performed in the radio domain, taking place in the era before white-light CME observations. Radio imaging of moving sources of synchrotron emission (known as a moving type IV radio burst) first provided a field strength diagnostic of 0.8\,G at a height\footnote{`Height' here means heliocentric distance e.g., solar surface is 1\,R$_{\odot}$} of 2\,R$_{\odot}$ \citep{boischot1968}. The analysis of moving type IVs lead authors to propose that these radio sources are from energetic electrons trapped in the magnetic field of ejected plasmoids in the corona \citep{dulk1971, smerd1971, riddle1980}. While these studies mainly concentrated on source morphology, kinematics and associated flare, some studies analysed the emission process in detail, identifying Razin suppressed gyrosynchrotron emission, allowing a magnetic field diagnostic of 6\,G at a height of 2\,R$_{\odot}$ \citep{bhonsle1980}. Studies during this era showed that the emission process for type IVs can be (gyro-)synchrotron in nature, although \citet{duncan1980} showed it can also be due to plasma emission. This highlighted that moving type IVs can provide a variety of diagnostics of the erupting plasmoid, either density or magnetic field diagnostics, depending on the emission mechanism.

Although these studies initially concluded that type IV radio bursts belonged to some form of ejected material, it was later realised that these radio bursts were associated with the newly discovered white-light `coronal transients' (or CMEs) \citep{kosugi1976}. \citet{gopal1987} identified a moving type IV burst in association with a CME to be from gyrosynchrotron radiation produced by $>$350\,keV electrons in a 2\,G magnetic field at 2.3\,R$_{\odot}$. \citet{stewart1982} and \citet{gary1985} also showed a CME to be closely associated with a moving type IV burst produced from plasma emission. The former study equated thermal to magnetic energy to estimate CME magnetic field strengths of $>$0.6\,G at a height of 2.5\,R$_{\odot}$. 

Perhaps the most famous case of a radio source associated with a CME was during the SOL1998-04-20 event \citep{bastian2001}. Observed by the Nan\c{c}ay Radioheliograph \citep[NRH;][]{kerdraon1997} at 164\,MHz, the flux density spectrum of this `radio CME' allowed the authors to conclude that this emission process was Razin-surpressed synchrotron radiation from 0.5-5\,MeV electrons in a CME magnetic field of $\sim$0.3--1.5\,G at a height of 3--4\,R$_{\odot}$. A similar case of a radio CME was reported in both \citet{maia2007} and \citet{demoulin2012}, 
with the former deriving a field strength on the order of $\sim$0.1--1\,G at $\sim$2\,R$_{\odot}$. The most recent observations have corroborated these findings, showing that gyrosynchrotron sources (type IV bursts) can be associated with a CME core, giving a field strength diagnostic of 1.4--2.2\,G at $\sim$1.9--2.2\,R$_{\odot}$ \citep{raja2014}. \citet{bain2014} studied a type IV source in a CME core finding a field strength of $\sim$3--5\,G at $\sim$1.5\,R$_{\odot}$, while \citet{tun2013} found a field strength as high as 5--15\,G for the same event. The discrepancy between the two results is possibly due to the different electron energy ranges and spectral slopes assumed in each analysis.

It is clear from the above studies that moving type IVs can be used as a useful diagnostic of CME magnetic field strength. However, moving type IVs are a rare phenomenon, with only about 5\% of CMEs being associated with such a radio burst \citep{gergley1986}. And amongst the many tens of thousands of CMEs observed since their discovery, the above studies represent relatively few events that have provided a means to estimate CME magnetic field strength. Despite the lack of observational studies of CME magnetic field, theoretical investigations have concluded that the magnetic field is both the trigger and driver of the eruption.
The models describe CME eruption using the free energy in a complex non-potential magnetic field, usually in the form of a flux rope \citep{aulanier2010, zuccarello2014}. The magnetic forces acting on this flux rope, whether expressed in the form of toroidal instability, magnetic pressures and tensions or Lorentz forces, are ultimately responsible for the eruption \citep[see][for a review]{chen2011}. This highlights the importance and need for further observations of CME magnetic field strength, yet it remains one of its most elusive properties. 

In our observations we report on another rare case of magnetic field measurement from non-thermal {gyrosynchrotron} radiation from a CME. {\color{black}We also highlight that at the same plane-of-sky (POS) position we observe plasma radiation as well as a source of soft and hard X-rays. Such a rare set of observations allows us to explore the relationship between radio and X-ray emitting electrons associated within the CME, ultimately allowing the eruption non-thermal electron properties and the magnetic field strength to be calculated.} 
In Section 2 we describe observations, in Section 3 we describe methods, including flux density measurements from NRH and Radio Solar Telescope Network \citep[RSTN;][]{guidice1979}, in Section 4 we discuss how these are used to obtain magnetic field measurements and in Section 5 we discuss the results in the context of CME plasma properties and conclude.


\section{Observations}
     \label{S-Observations} 
     
\begin{figure}[!t]
    \begin{center}
    \includegraphics[scale=0.32, trim=1cm 5.5cm 0cm -0.2cm]{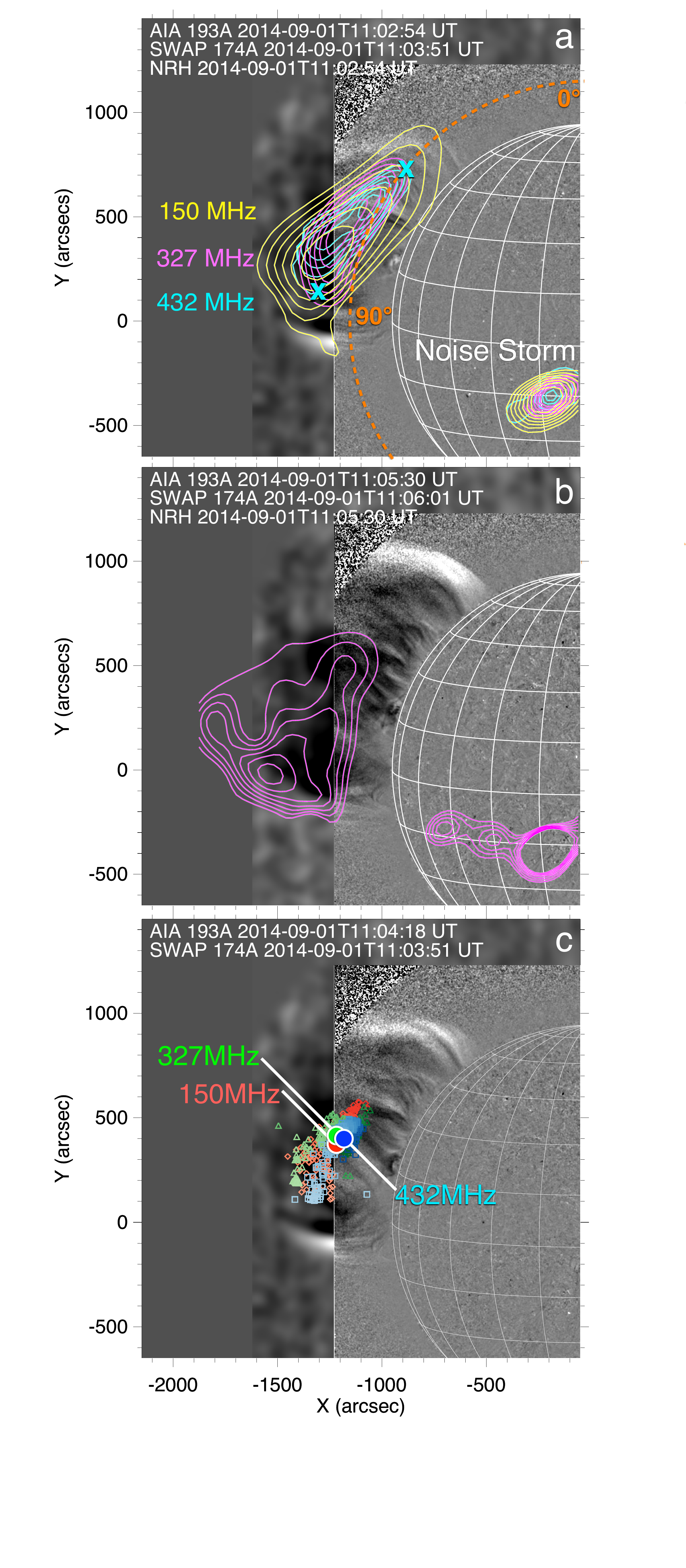}
    \caption{(a) Eruption from the east limb observed using an AIA 193\,\AA~and SWAP 174\,\AA~ratio images, superimposed with NRH 150, 327 and 432\,MHz contours. By 11:01\,UT an EUV `bubble' can be seen developing from the east limb. At the same location in the plane of sky, large radio sources appear at all NRH frequencies (just three are shown here). The sources remain at this position until $\sim$11:05\,UT. The radio source on disk is a type I noise storm, unrelated to the event in question. (b) At 11:05:30\,UT, the lower frequency sources (150\,MHz) decrease in intensity and the remaining higher frequencies are situated to the southern flank of the eruption (just 327\,MHz is shown here). At this time a strong and sharply defined EUV front can be seen propagating toward the north pole.  The noise storm on disk is elongated due to the side-lobes of the telescope beam. (c) Location of sources through time, colors indicate frequency and dark-to-light shading in colour indicates progression through time. The indicated circles are the centroids of each frequency cluster separately.}
    \label{fig:aia_nrh}
    \end{center}
\end{figure}     
     
The SOL2014-09-01 event was associated with a flare occurring 36$^{\circ}$ beyond the east solar limb at N14E126, observed by the Extreme Ultraviolet Imager \citep[EUVI;][]{wuesler2004} on board the the {\it Solar Terrestrial Relations Observatory} Behind spacecraft, with an estimated GOES class of X2.4 \citep{ackermann2017}. Given this was a behind the limb flare, no increase in X-ray flux was recorded by the GOES spacecraft, see Figure~\ref{fig:goes_dam_orfees}. The event was associated with a fast CME with a speed of $\sim$2000\,km\,s$^{-1}$, first appearing in the Large Angle Spectroscopic Coronagraph \citep[LASCO;][]{bru95} (C2) at 11:12\,UT (see \citet{pesce-rollins2015, ackermann2017} for a description of the latter part of this event not studied here, including gamma ray observations with \emph{Fermi}-LAT).

Beginning at 11:00\,UT, a variety of solar radio bursts were observed by the Nan\c{c}ay Decametric Array \citep[NDA;][]{lecacheux2000} and the Orf\'{e}es spectrograph between 10--1000\,MHz, see Figure~\ref{fig:goes_dam_orfees}. The radio event begins with a type II radio burst at 11:00\,UT at $\sim$40\,MHz in the NDA spectrograph, followed by a complex and bursty emission which lasts for $\sim$30\,minutes. In the Orf\'{e}{e}s frequency range between 11:01--11:06\,UT, we observe bursty emission extending up to $\sim$$200-300$\,MHz, with a faint, smooth and broadband emission at higher frequencies which we label here as a type IV radio burst. We concentrate on this radio burst for the remainder of this paper.
\begin{figure}[!t]
    \begin{center}
    \includegraphics[scale=0.25, trim=2cm 3cm 0cm 0cm]{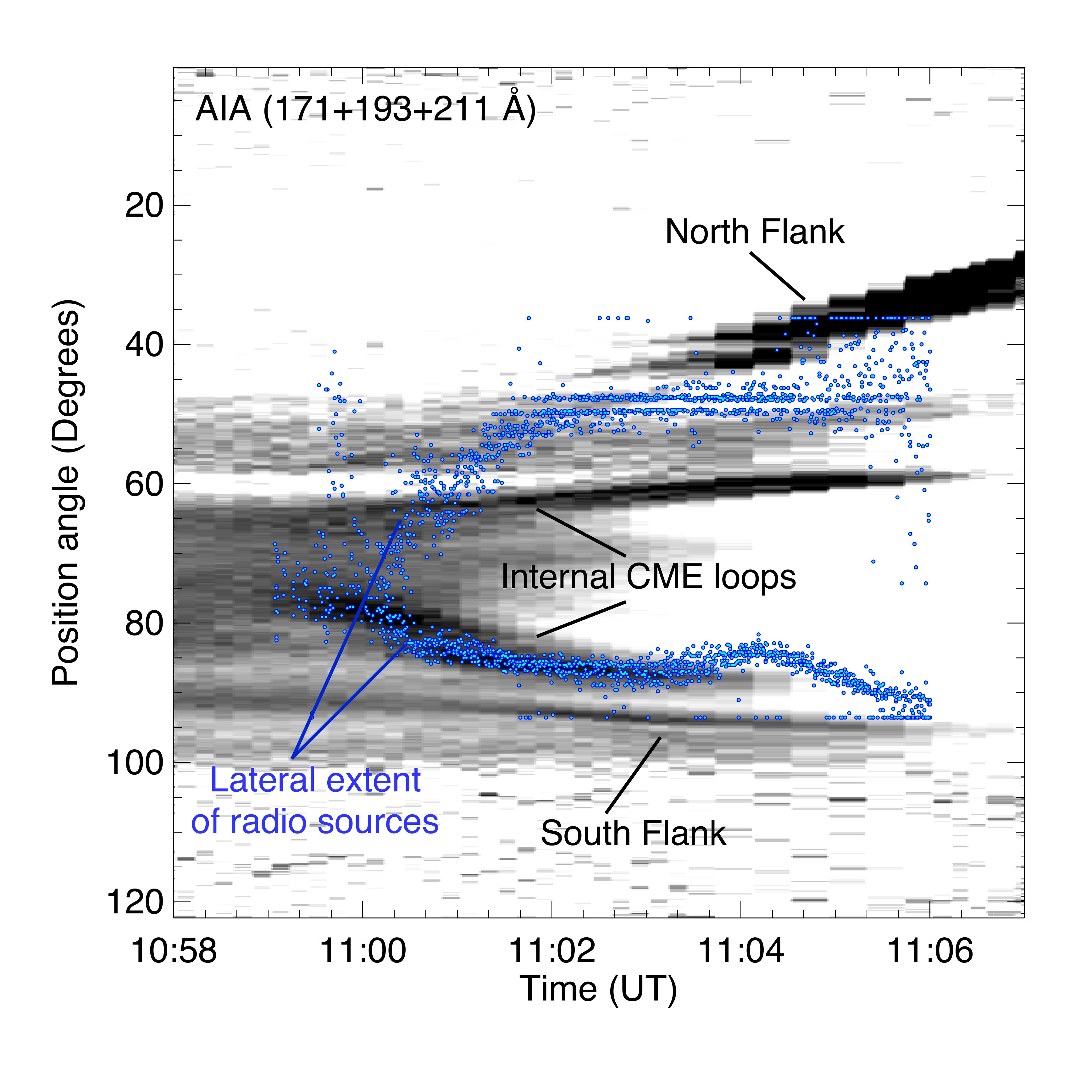}
    \caption{{Distance-time map (reverse colour) produced from intensity traces extracted from the orange circle in Figure~\ref{fig:aia_nrh}(a). The map is produced from a normalisation then addition of AIA 211\,\AA, AIA 193\,\AA, and AIA 171\,\AA~passbands. The upper and lower CME flanks are indicated, along with the expanding inner loops of the CME. The blue points mark the position angles of the upper and lower lateral extent of the 408, 445 and 432\,MHz radio sources -- the dark blue `$\times$' symbols in Figure~\ref{fig:aia_nrh}(a) mark these two points on the 432\,MHz source for a single time. The radio source expansion follows the CME expansion quite closely. The southern extent of the radio sources in particular follow the expansion of the CME southern internal loop.}}
    \label{fig:cme_radio_expansion}
    \end{center}
\end{figure}     
     
At the time of the type IV burst, an eruption can be seen developing from the east limb using the Atmospheric Imaging Assembly \citep[AIA;][]{lemen2012}~193\,\AA~filter and the Sun Watcher using Active Pixel \citep[SWAP;][]{berghmans2006} 174\,\AA~passsband as shown in Figure~\ref{fig:aia_nrh}. The eruption is first seen as disturbed loops beginning to emerge at $\sim$10:59\,UT which then develop into an EUV `bubble' with a strong and sharply defined EUV wave propagating toward the north pole, a snapshot of which is shown in Figure 2(b).

At the same location as the eruption, we observe large radio sources using multiple frequencies of the Nan\c{c}ay Radioheliograph \citep[NRH;][]{kerdraon1997}, see Figure~\ref{fig:aia_nrh}(a). The NRH contours are 150, 327 and 432\,MHz scaled between 50\% and 100\% of the maximum brightness temperature for each source individually. Initially, at 150 MHz the sources have a full-width-half-maximum (FWHM) in the southeast-northwest direction of $\sim$0.7\,R$_{\odot}$. All other NRH frequencies show a large source at a similar location, with the source having FWHM of $\sim$0.5\,R$_{\odot}$ at 327 MHz and reducing to $\sim$0.45\,R$_{\odot}$ at 408\,MHz and above. After 11:05\,UT, the lower frequency sources (150\,MHz) have disappeared, while the high frequency sources move to the southern flank of the eruption, as seen in Figure~\ref{fig:aia_nrh}(b) -- just 327 MHz is shown for simplicity, the higher frequency sources are smaller but at a similar position.

\begin{figure}[!b]
    \begin{center}
    \includegraphics[scale=0.45, trim=0cm 1cm 0cm 1cm]{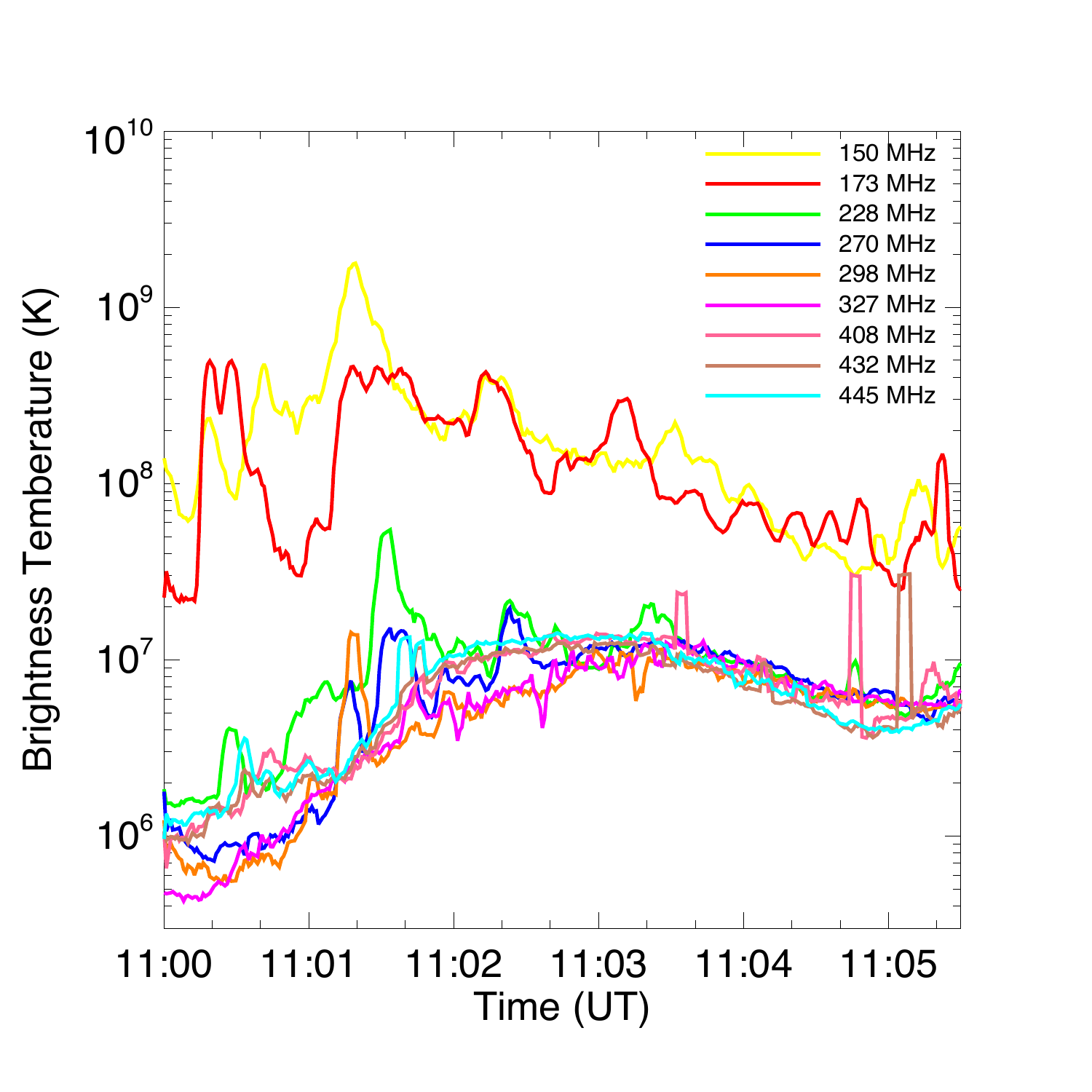}
    \caption{Maximum brightness temperature as a function of time for the radio sources at all NRH frequencies. The 150\,MHz and 173\,MHz sources show much higher brightness temperatures than 228\,MHz and above.}
    \label{fig:brightness_temp}
    \end{center}
\end{figure}    

Figure~\ref{fig:aia_nrh}(c) shows the positions of the radio source maxima at 150 (red), 327 (green) and 432 (blue)\,MHz over time between 11:01\,UT and 11:05:30\,UT, overlaid on an AIA193\,\AA~running ratio image at 11:02:01\,UT. The shading of the points from dark to light represents change in position with time. All of the points are generally clustered around the same area at the center of the eruption. Each source shows a consistent progression southwards at a speed of $\sim$1500\,km\,s$^{-1}$, which is close to the speed of the CME southern flank of $\sim$1200\,km\,s$^{-1}$ at an altitude of $\sim$0.2\,R$_\odot$ in the southerly direction. 
{A closer study of the relationship between CME and radio source expansion is shown in Figure~\ref{fig:cme_radio_expansion}. This is a distance-time (dt) map produced from intensity traces taken from 171, 193, and 211\,\AA~passabands along the orange circle of fixed radius of $\sim$1.2\,R$_{\odot}$ in Figure~\ref{fig:aia_nrh}(a). Each intensity trace was normalised in brightness and summed across the three passbands. The upper and lower CME flanks are indicated on the dt-map, along with the expanding inner loops of the CME. The dark blue points mark the position angles of the upper and lower lateral extent of the 408, 445 and 432\,MHz radio sources. For example, the two blue `$\times$' symbols mark the upper and lower lateral extent of the 432 MHz source for a single time in Figure~\ref{fig:aia_nrh}(a). Figure~\ref{fig:cme_radio_expansion} shows that both the CME and radio sources have a common lateral broadening. This is particularly noticeable in the expansion of the southern internal loop of the CME, which is followed closely by the southern extent of the radio sources e.g. the radio source expands to the south at the same rate as the internal CME loops. This provides good evidence that the radio sources belong to these internal loops, possibly in the core of the CME. We discuss further the possible locations of the radio sources in Section~\ref{S-lifetime-location}.}

The indicated circles in Figure~\ref{fig:aia_nrh}(c) demarcate the average positions of an individual set of frequency points, showing that the 150, 327 and 432\,MHz sources are closely spaced (all NRH frequencies from 228--445\,MHz are clustered around this position). The majority of frequencies being located at the same position is an indicator that the origin of at least some of these sources is not plasma emission i.e., 
with plasma emission we expected to see some stratification in frequency, due to the stratification in density of the environment from which the emission comes. If sources generally have the same position, there is a possibility they may be from a gyrosynchrotron source. To better distinguish the nature of the radio emission sources and emission mechanisms, 
in the following we investigate both the brightness temperatures and flux densities as a function of time and frequency, employing the use of both NRH and the RSTN.
     
  \begin{figure*}[!t]
    \begin{center}
    \includegraphics[scale=0.6, trim=0.5cm 9cm 0cm 0cm]{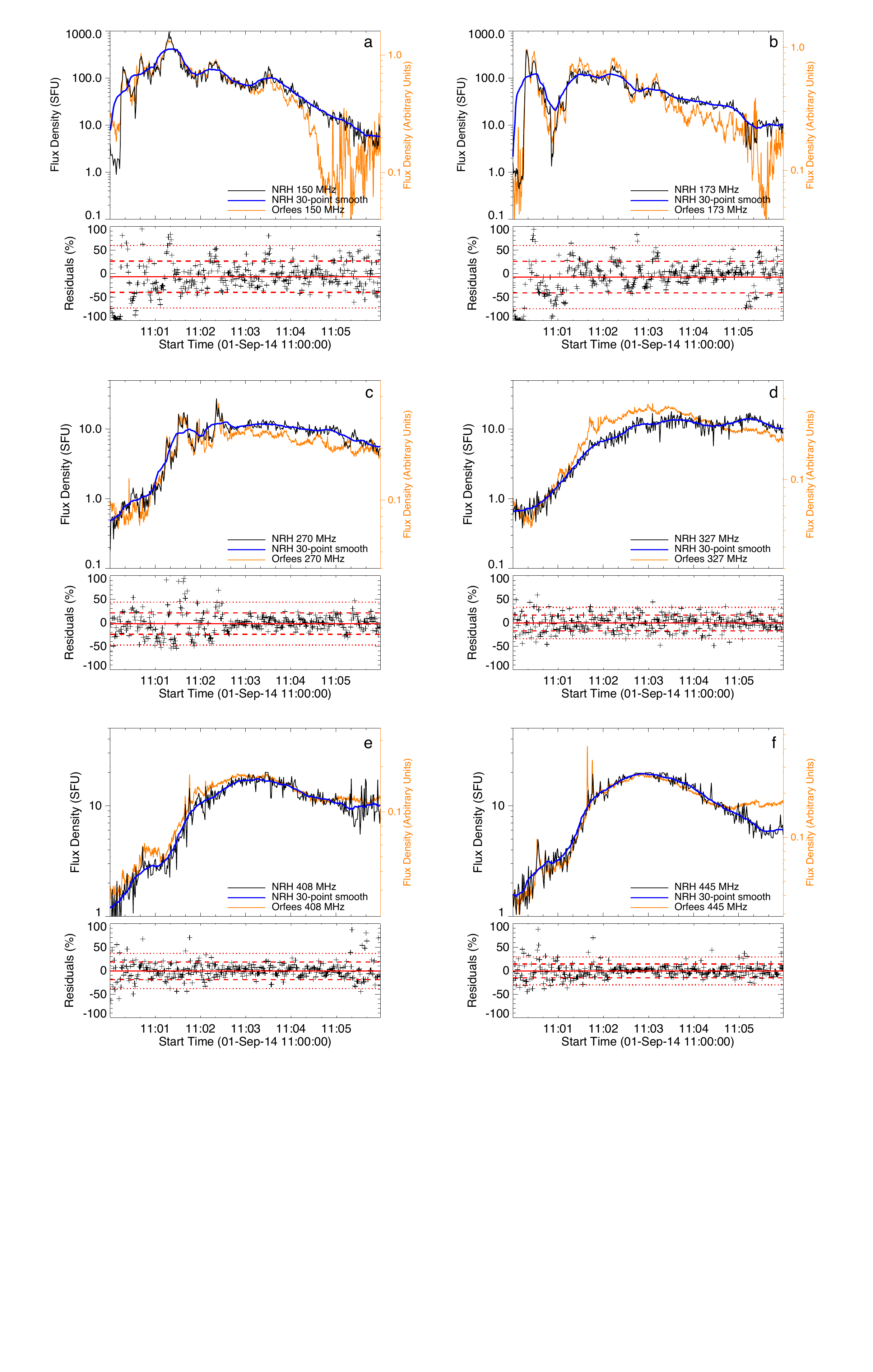}
    \caption{Flux densities of the large radio source for all NRH frequencies (black; units of SFU) and flux density from Orf\'{e}es at the same frequency (orange; arbitrary units). The blue line is a 30-point box-car smooth of the NRH time profiles. The difference between the black and blue curves provide the residuals over time for each panel. Residuals larger than 2$\sigma$ are indicative of sporadic radio bursts. The 150--173\,MHz sources have much larger flux densities and are also more bursty in nature (larger residual spread). At 228\,MHz and above the flux densities are much weaker, while at 300\,MHz and above the flux is weaker and much more smoothly varying in time i.e. the residuals mainly varying by less than 2$\sigma$, indicative of flux variation due Gaussian noise only (very few sporadic radio bursts) above these frequencies.}
    \label{fig:nrh_orfees_flux}
    \end{center}
\end{figure*}        
     
\subsection{Brightness Temperature: NRH Observations}     
The peak brightness temperature of the radio sources as a function of time is shown in Figure~\ref{fig:brightness_temp}. It shows that the 150 and 173\,MHz sources peak in brightness temperature at 1.5$\times$10$^9$\,K and 5$\times$10$^8$\,K, respectively. Frequencies above 228\,MHz show brightness temperatures over an order of magnitude lower than this, with most peaking at 1$\times$10$^7$\,K. Also, similar to the dynamic spectrum, the variation of brightness temperature in time of the 150 and 173\,MHz sources is more bursty and sporadic than the smoothly increasing and decreasing time profiles of the higher frequency sources. 
This difference in brightness temperature variation and time profile, as well as the association of the high frequency sources with the much smoother type IV in Orf\'{e}es, is suggestive of the different emission mechanisms between low frequency and higher frequency sources i.e., low frequencies ($<$170\,MHz) are bursty and intense while higher frequencies are smoothly varying and weaker. 


\subsection{Flux Density: NRH Observations}
To further investigate the difference between high and low frequencies in NRH, we calculate the flux density of the sources as a function of time and compare them directly to the Orf\'{e}es flux (arbitrary units). 
The results are shown in Figure~\ref{fig:nrh_orfees_flux}, with NRH flux density profiles shown in black and the Orf\'{e}es flux profiles in orange. At each frequency the NRH and Orf\'{e}es flux density profiles are directly comparable, showing that the sources imaged off the east limb were responsible for the radio bursts in the spectrogram.
As expected, the flux density of the 150 and 173\,MHz reach higher values, at well above 300\,SFU and are almost 2 orders of magnitude larger than the remaining NRH frequencies e.g., at 270\,MHz and above, the flux density peaks just above 10\,SFU.

The flux density profiles over time, particularly above 270\,MHz, appear to be composed of roughly two components; the smoothly varying rise and fall of flux density over $\sim$5\,minutes (with peak at $\sim$11:03\,UT) and more sporadic and sharply defined bursts which appear on timescales of seconds. Sporadic bursts amongst the smoothly varying profiles may be either fluctuation due to statistical noise or an actual signal from a short time interval radio burst. In order to distinguish between the two possibilities, we perform a 30-point boxcar smooth on the flux profile (shown in blue) and subtract this from the original unaltered profile (black). This allows us to examine the residuals due to this subtraction i.e., the size of the short-duration bursts as a function of time.

The residuals are shown in the panels below each flux density profile. 
The red solid line represents the mean residual values, while the dashed and dotted lines are $\pm$1 and 2 standard deviations ($\sigma$), respectively. For the low frequencies, as expected, the residuals show a large spread in values and frequently show bursts larger than 2$\sigma$, due to the bursty nature of the plasma emission at these frequencies. At frequencies of 327\,MHz and above the residuals have a narrower range, showing only brief intervals of bursty emission above 2$\sigma$; any short duration variation is mainly due to statistical noise only. Hence in the following analysis we consider the NRH frequencies at 327\,MHz and above to be observation of the smoothly varying radiation (gyrosynchrotron, as will be shown) only.

\begin{figure}[!t]
    \begin{center}
    \includegraphics[scale=0.30, trim=1.0cm 7cm 0cm 1.0cm]{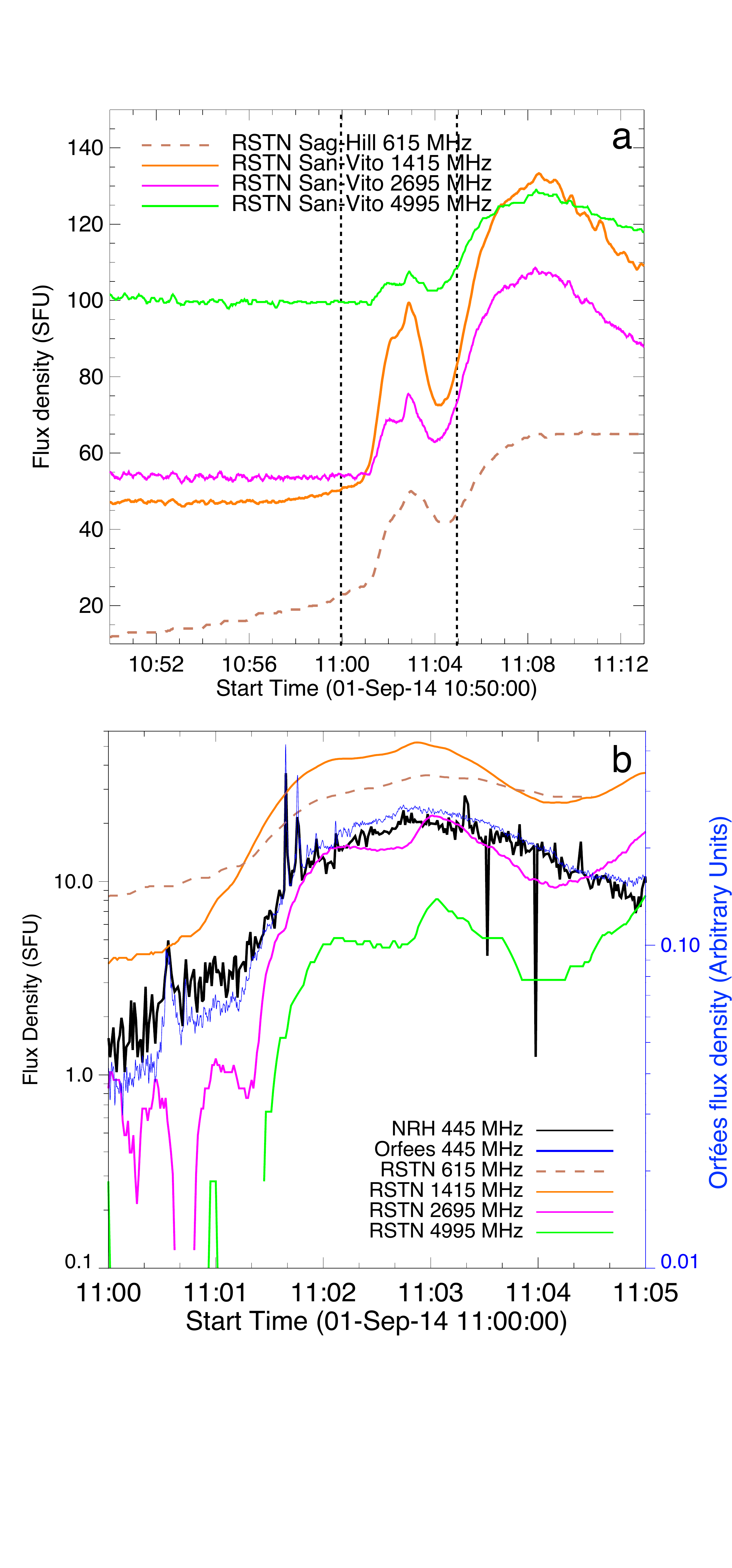}
    \caption{(a) Flux densities at 615\,MHz from Sagamore Hill site and 1.4, 2.7 and 5.0\,GHz from the San-Vito site of the RSTN network. Each of the flux density curves has a steady background which lasts for several hours prior to the event. We use this to perform a background subtraction on each frequency {(the rise in flux of 615 and 1415\,MHz before the time of interest is taken into account in the flux uncertainty at these frequencies in the analysis below)}. The dashed lines represent the region of interest (the type IV burst). (b) Zoom of the RSTN flux density spectrum (between the dashed lines demarcated in (a)) with the NRH and Orf\'{e}es 445\,MHz flux density curve for comparison. The flux densities from all three instruments show a good relationship of flux density variation with time.}
    \label{fig:rstn_flux}
    \end{center}
\end{figure}

\subsection{Flux Density: RSTN Observations}

Above 600\,MHz we use flux density observations of the San Vito and Sagamore Hill sites of the RSTN network. The San Vito observatory recorded increase in flux at three separate frequencies of 1.4, 2.7 and 5.0 GHz, as shown in Figure~\ref{fig:rstn_flux}(a). Sagamore Hill provided a flux density measurement at the same frequencies\footnote{ The fluxes from the different RSTN sites generally matched to within 15--20\%; this shows that the flux calibration from each site is reasonable and fluxes amongst the common frequencies between the two sites were averaged } and a extra flux density measurement at 610\,MHz. 

Our time of interest is between the vertical dashed lines in Figure~\ref{fig:rstn_flux}(a). The smooth rise and fall is again observed between 11:00--11:05\,UT, after which the flux density starts to rise again to larger values (this rise in flux is from a second part of the event not associated with the type IV in question; for discussion on this second part of the event see \citet{ackermann2017}). Throughout the day, RSTN recorded a steady background prior to the event. This enabled a background subtraction (mean flux density between 10:30--10:50\,UT) such that the only flux density increase is from our radio sources of interest. {In the 615\,MHz and 1415\,MHz there is a rise in flux above the background before our time of interest. This initial rise could be the start of flux from the type IV, but may also be the beginning of the radio emission which peaks after our time of interest (after 11:05\,UT). Due to this ambiguity, the flux density uncertainties at 615 and 1415\,MHz include this extra flux rise of $\sim$13\,SFU and $\sim$6\,SFU, respectively, in the analysis below.}

The background-subtracted RSTN flux density profiles are shown in Figure~\ref{fig:rstn_flux}(b), and are directly comparable to the NRH and Orf\'{e}es profiles. Together, NRH and RSTN then provide measurement of the flux density from 150\,MHz to 5\,GHz, allowing the construction of a flux density spectrum.

%
%
\begin{figure}[!ht]
    \begin{center}
    \includegraphics[scale=0.35, trim=1.5cm 6.5cm 0cm 0cm]{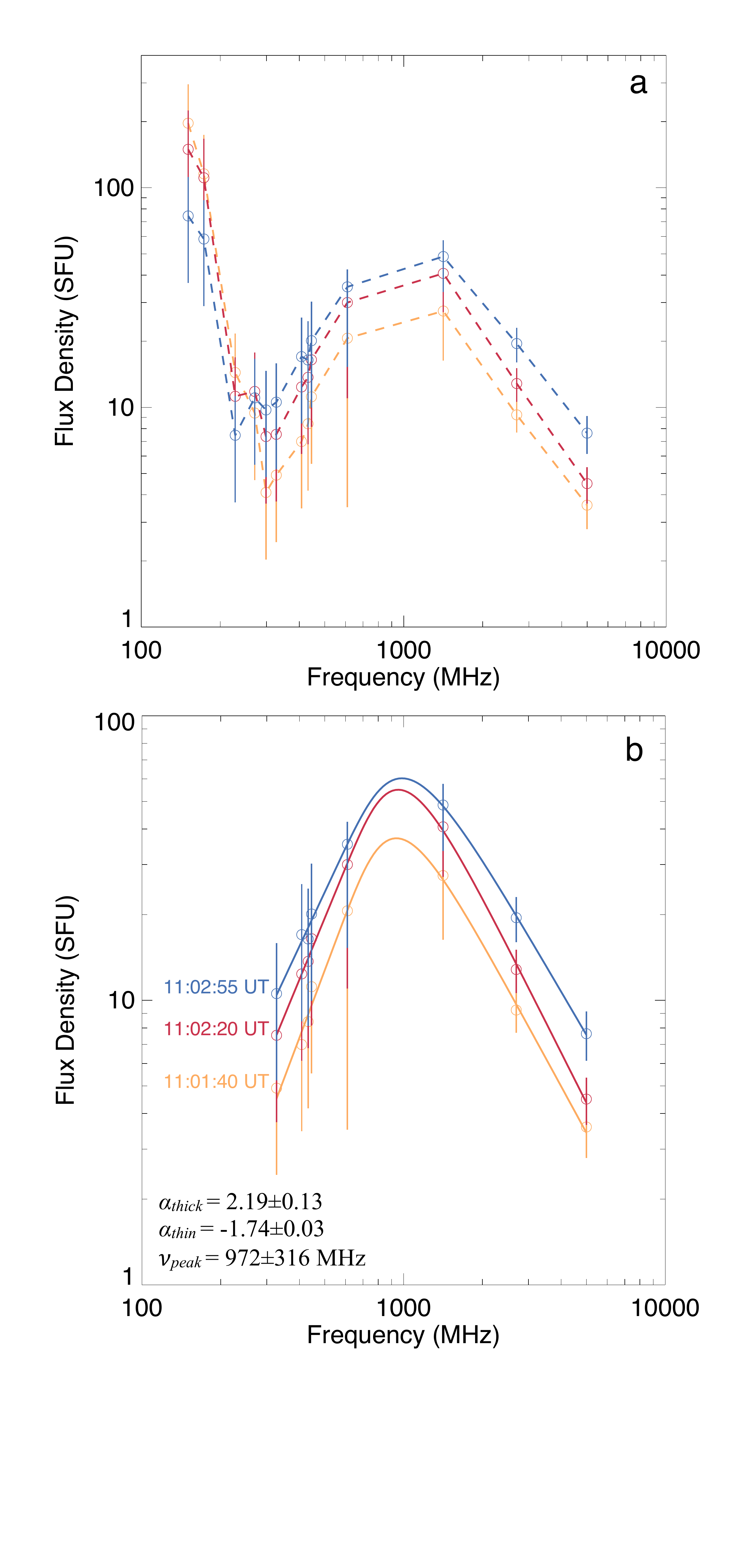}
    \caption{(a) Flux density spectrum of NRH data below 445\,MHz and RSTN data above 615\,MHz, at three separate times between 11:01:40\,UT to 11:03:10\,UT (times indicated in bottom panel). The flux densities are high at 150 and 173\,MHz, fall to a minimum at $\sim$300\,MHz, reach a peak $\sim$1\,GHz and fall again. This is characteristic of a gyrosynchrotron spectrum at frequencies higher than $\sim$300\,MHz with plasma emission dominating at lower frequencies. (b) A parametric fit of a gyrosynchrotron spectrum from 298\,MHz to 4.5\,GHz at three separate times.}
    \label{fig:flux_density_fit}
    \end{center}
\end{figure}

\subsection{Flux Density Spectrum}

Figure~\ref{fig:flux_density_fit}(a) shows the NRH and RSTN flux density values as a function of frequency for three different time intervals. The NRH error bars are from the 2$\sigma$ values of the residuals at each frequency as a measure of statistical variation in the signal, as described above. We have added to this a 20\% absolute calibration uncertainty on the NRH flux density values. The RSTN uncertainty is set to 21\%, which takes into account a 20\% flux density calibration uncertainty plus 1\% variation in the background due to statistical noise. {Note that the 615 and 1415\,MHz negative uncertainty bars are larger because they take into account the initial rise in background flux before 11:00\,UT at these frequencies, as stated above.}

From Figure~\ref{fig:flux_density_fit}(a), the flux density spectrum is high at low frequencies, falls to a minimum at $\sim$300\,MHz and reaches a maximum again at $\sim$1\,GHz. Such a spectrum is characteristic of previous results showing plasma emission at low frequencies and gyrosynchrotron emission at larger frequencies \citep{dauphin2005}, and is described in detail in \citet{nita2002}. This corroborates the analysis of the flux density time-profile residuals above i.e., below 300\,MHz we are likely observing a mixture of bursty plasma emission and gyrosynchrotron emission, while at higher frequencies we observe only gyrosynchrotron emission.

%
%
  
\section{Calculating magnetic field in the radio source}

The construction of a flux density spectrum of the type IV burst using NRH and RSTN from 300--5000\,MHz allows us to estimate the CME magnetic field strength in the radio source using two methods; firstly, we use the approximations to gyrosynchrotron radiation theory from \citet{dulk1982}. {\color{black}This analysis requires a calculation of non-thermal electron properties from the coronal X-ray emission at the time of our event. This leads to useful comparisons of the X-ray and radio emitting electron populations associated with the CME.} In the second method we use a full gyrosynchrotron numerical model \citep{simoes2006} to estimate a more accurate value for the CME magnetic field strength as well as the characteristics of the non-thermal electron distribution responsible for the radio emission.

\subsection{Magnetic field estimate from the Dulk \& Marsh approximations}     
	\label{S:DM_approx}

As outlined above, it is only at 300\,MHz and above that we observe gyrosynchrotron emission. We therefore choose data points above this frequency to which a parametric fit of a gyrosynchrotron spectrum is to be applied. Gyrosynchrotron flux density spectra can be approximated with a generic parametric equation in the form of
\begin{equation}
S_{\nu} = S_{peak}\left(\frac{\nu}{\nu_{peak}} \right)^{\alpha_{thick}} \left \{ 1 - exp\left [ -\left(\frac{\nu}{\nu_{peak}} \right)^{\alpha_{thin}-\alpha_{thick}} \right ]  \right \}
\label{eqn:flux_density_fit}
\end{equation}    
 \citep{StahliGaryHurford:1989} where $S_{\nu}$ is flux density as a function of frequency $\nu$, {$S_{peak}$ is the peak flux density, $\nu_{peak}$ is the frequency at which the peak flux density occurs}. {For $\nu>>\nu_{peak}$, the expression reduces to $S_{\nu} = S_{peak}\left(\nu/\nu_{peak} \right)^{\alpha_{thin}}$, where $\alpha_{thin}$ is the spectral index on the optically thin side of the spectrum (the negative slope). For $\nu<<\nu_{peak}$ it reduces to $S_{\nu} = S_{peak}\left(\nu/\nu_{peak} \right)^{\alpha_{thick}}$, where $\alpha_{thick}$ is the spectral index on the optically thick side of the spectrum (positive slope). Although the expression is parametric, it takes into account the general behaviour of gyrosynchrotron spectra in optically thin and thick regimes.}
 
The equation was fit to our data for flux density spectra every second between 11:01:40\,UT --11:03:10\,UT (the time during which RSTN reaches flux values above background, see Figure~\ref{fig:rstn_flux}); three spectra throughout this interval are shown in Figure~\ref{fig:flux_density_fit}(b). The resulting fits give average spectral indices of $\alpha_{thick}=2.19\pm0.13$ and $\alpha_{thin}=-1.74\pm0.03$. These spectral indices fall close to the range of those expected for gyrosynchrotron emission i.e., $\alpha_{thick}=2.9\pm0.1$ and range of $\alpha_{thin}=[-4,-1.5]$  \citep{dulk1973, dulk1985}. 
Our results also agree with a more recent statistical study of gyrosynchrotron spectral slopes of $\alpha_{thin}=-2.51^{+0.75}_{-0.90}$ and $\alpha_{thick}=1.79^{+1.04}_{-0.53}$ \citep{nita2004}. Also, the average peak frequency we find here is $\nu_{peak}=972\pm316$\,MHz, which is smaller than the median peak frequency of 6.6\,GHz in the \citet{nita2004} study\footnote{However the authors noted that a small subset of radio events (only 5\% of 412 events) had peak frequencies below 1.2\,GHz. Our event then seems to belong to this rare subset.}.

Now, from \citet{dulk1982} it is possible to make empirical approximations which relate the peak frequency in the flux density spectrum to the magnetic field via
\begin{equation}
\nu_{peak} \approx 2.72 \times 10^{3} 10^{0.27\delta} (\sin\theta)^{0.41+0.03\delta}(NL)^{0.32-0.03\delta} \times B^{0.68+0.03\delta}
 \label{E-nu_peak}    
\end{equation}
where $\theta$ is the angle between the radiation $k$ vector and the magnetic field direction, $N$ is the number density of non-thermal electrons, $L$ is the length of the radiating region along the line of sight and $B$ is the magnetic field; $\delta$ is the spectral index of the electrons in the power-law distribution and is related to the optically thin spectral index of the flux density spectrum by
\begin{equation}
\delta = |-1.1(\alpha_{thin}-1.2)|
\label{eqn:alpha_thin}
\end{equation}
as given in \citet{dulk1982}; in our case this results in $\delta_{radio}=3.2\pm0.3$ (note this is positive, but represents a negative slope in the spectrum; we use subscript `radio' here to distinguish it from the same parameter derived from X-ray below). We may therefore use our value for peak frequency and $\delta_{radio}$ to calculate the magnetic field strength, provided we know $\theta$, $L$ and $N$. Firstly, we assume a complicated magnetic structure along the line of site, which would give an average viewing angle to the magnetic field of $\theta$$\sim$45$^{\circ}$; such a complicated structure would also lead to the low levels of polarisation ($<$5$^{\circ}$) that we see in NRH. For $L$ we take the emission region along the line of site to be no greater than the average source size in the plane of the sky between 327--445\,MHz in NRH ($L$$\sim$0.45\,R$_{\odot}$). Perhaps the most uncertain of these properties is $N$, which cannot be estimated directly from the radio data. However, we may derive an estimate of $N$ using X-ray observations from \emph{Fermi} Gamma-ray Burst Monitor \citep[GBM]{MeeganLichtiBhat:2009} at the time of the event. We may only use such a value for $N$ provided we can show the population of electrons emitting X-ray is the same or has a relationship with the population emitting radio. In the following we test for such a relationship.

\subsubsection{Energetic electron properties deduced from radio and X-ray observations}

\begin{figure}[!h]
    \begin{center}
    \includegraphics[scale=0.3, trim=1.5cm 1cm 0cm 0cm]{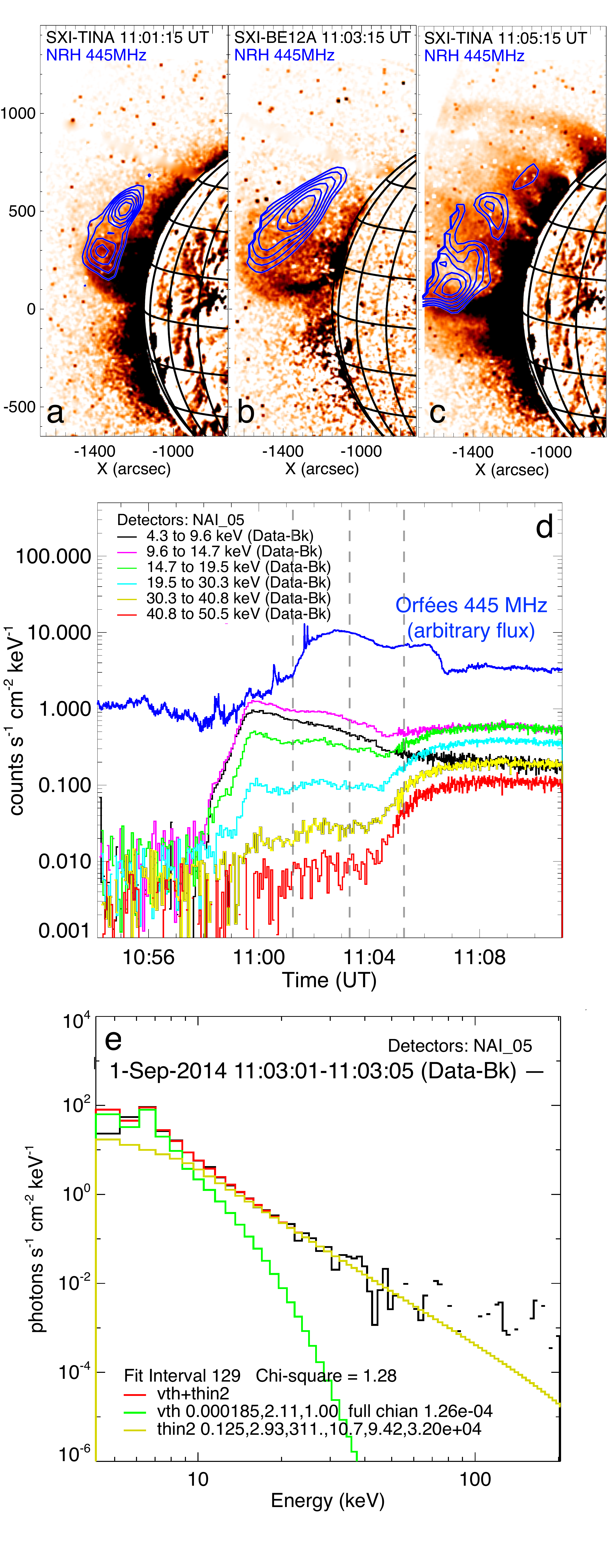}
    \caption{ (a)-(c) GOES SXI images showing the evolution of the eruption in soft X-rays, with NRH 445 MHz contours for comparison. (d) Fermi GBM counts from 4.2--50.5\,keV channels. An increase in counts can be seen in the channels between 4.2--40.8\,keV at the time of the event. A 445 MHz flux density curve (with arbitrary flux) from Orf\'{e}es is shown in blue for comparison. Although the radio and X-ray start to rise at similar times, the radio peak is $\sim$3 minutes after the X-ray peak. The vertical dashed lines are times of the images above. (e) The photon energy spectrum fit with a thermal and thin-target model, given that that the flare was beyond the limb (assumption of no observed thick-target source).}
    \label{fig:fermi_gbm}
    \end{center}
\end{figure}

 {\color{black} The soft X-ray (SXR) emission for this event was imaged by GOES Soft X-ray Imager \citep[SXI;][]{lemen2004}, shown in Figure~\ref{fig:fermi_gbm}(a)-(c) overlayed with NRH 445 MHz contours from 50-100\% of max brightness temperature. {The eruption appears over the east limb at $\sim$11:01\,UT, with the radio emission concentrated in two lobes, the southerly of which is in the same region as the SXR emission. The extended SXR emission grows, with the radio emission covering mainly the upper half of the eruption. As the eruption develops the radio emission becomes more fragmented and concentrates towards the centre and south of the eruption. It is clear that at least part of the radio and SXR emission overlap, which indicates some relationship between the two.}
 
 The HXR activity was also observed by \emph{Fermi} GBM detector 5\footnote{See \cite{ackermann2017} for further X-ray observations, including Konus-\emph{WIND} 20-78\,keV flux.}, the counts of which begin to increase at 10:57\,UT in channels with energies from 4.3--50.5\,keV, reaching a peak at $\sim$10:59--11:00\,UT, see Figure~\ref{fig:fermi_gbm}(d). The Orf\'{e}{e}s 445\,MHz flux is shown with arbitrary flux scaling for comparison.}
The radio flux starts to rise at 10:58\,UT, but peaks approximately 3 minutes after the X-ray flux. However, at the time of the radio peak flux, a second small enhancement may be observed in the X-ray flux at $\sim$11:01--11:03:30\,UT, particularly noticeable in the 19--30\,keV channel (the differences in the X-ray and radio peaks are discussed in Section~\ref{S-number_dsicussion}). {The fact that the thermal SXR emission and radio emission show some spatial correspondence in Figure~\ref{fig:fermi_gbm}(a)-(c) and also show some temporal relationship in panel (d) lead us to investigate and compare the properties of X-ray and radio emitting CME electrons.}

Figure~\ref{fig:fermi_gbm}(e) shows the X-ray photon energy spectrum at $\sim$11:03\,UT (time of the second peak). Since the flare was $\sim$36$^{\circ}$ behind the limb, we assume that this photon spectrum is from a coronal thin-target X-ray source, similar in observation to \citet{kane1992, krucker2010, simoes2013}. We therefore fit the X-ray flux spectrum with a model for optically thin thermal component (green) combined with a thin-target bremsstrahlung radiation spectrum from an isotropic electron flux density distribution (non-thermal component; yellow). We produced this fit for all intervals between 11:01:00--11:03:30\,UT (around the second peak in the X-ray flux) and computed average values of all fitted parameters, resulting in mean thermal plasma temperature of 29\,MK and emission measure of $\xi=2.0\times10^{45}$\,cm$^{-3}$. 
The fit also gives a non-thermal low energy cut-off of $E_0\sim$9\,keV and an integrated electron flux of $[n_{0}V_0\overline{F}]=7.2\times10^{53}$\,electrons\,cm$^{-2}$\,s$^{-1}$\,keV$^{-1}$, where $n_0$ is the ambient electron density, $V_{0}$ is the thermal emitting volume, and $\overline{F}=\int FdE$ i.e., the integrated electron flux density over energy. 

\begin{figure}[!t]
    \begin{center}
    \includegraphics[scale=0.40, trim=1.5cm 0.5cm 0cm 0cm]{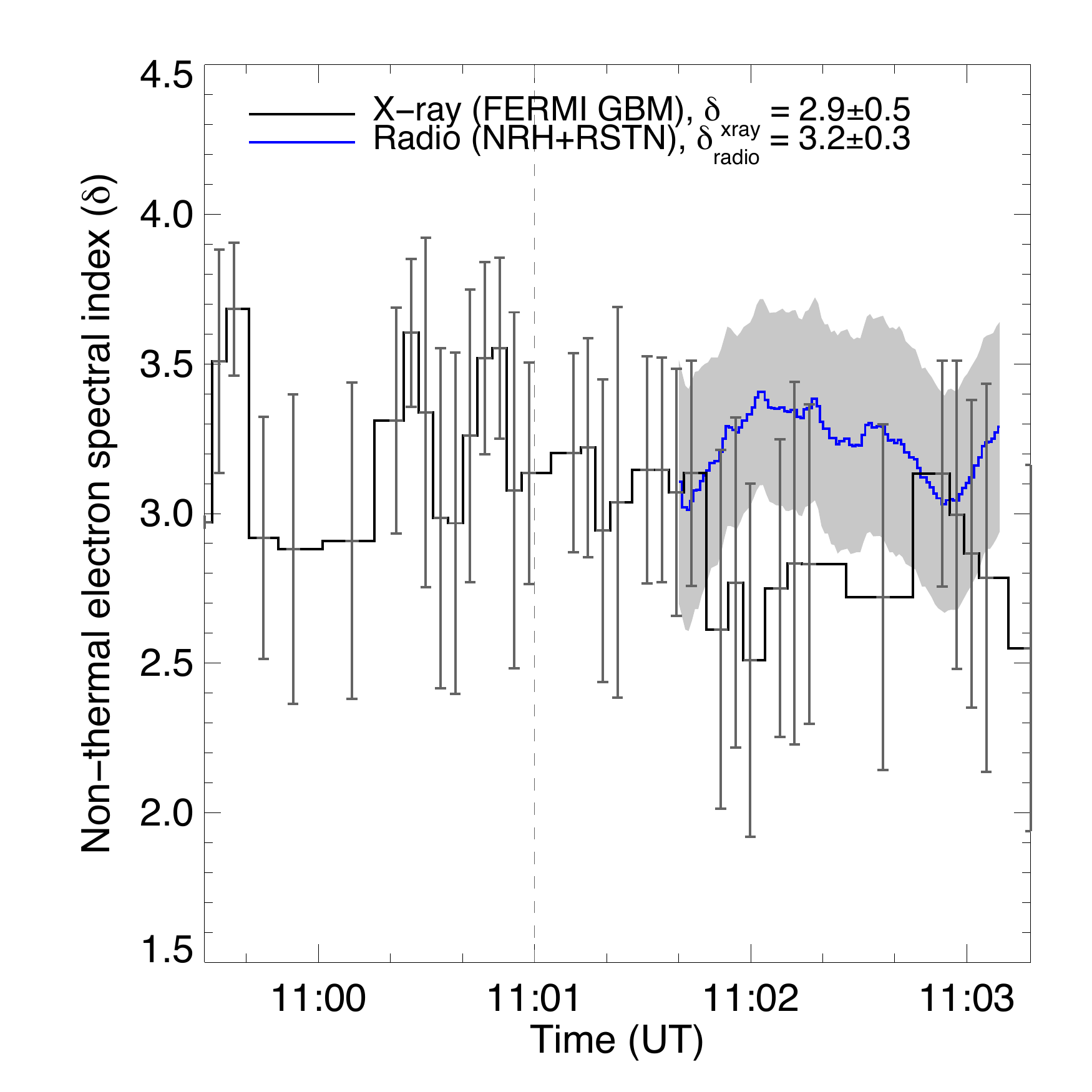}
    \caption{Non-thermal electron spectral indices derived from both X-ray energy spectra (black) and radio flux density spectra (blue) over time. The average values is taken during the time of the second X-ray peak after 11:01\,UT (after the dashed line). This gives mean values of  $\delta_{radio}=3.2\pm0.3$ and that from X-ray being $\delta_{xray}=2.9\pm0.5$. This is indicative of both radio and X-ray emission being produced from electron populations with similar characteristics e.g., the electrons producing thin-target X-ray emission may be closely related to those which produce radio.}
    \label{fig:spectral_indices}
    \end{center}
\end{figure}

Most interestingly, the non-thermal electron power law index derived using X-ray observation $\delta_{xray}$ (from the fit of the thin-target model) may be directly compared to that derived from radio $\delta_{radio}$; Figure~\ref{fig:spectral_indices} shows these spectral indices over time. They remain relatively constant, with the average values between 11:01:00--11:03:30\,UT (after the dashed line, around the second small peak in X-rays) being  $\delta_{xray}=2.9\pm0.5$ and radio resulting in $\delta_{radio}=3.2\pm0.3$. The similarity of these two spectral indices is evidence that both radio and thin-target X-rays came from a similar non-thermal electron population. Generally, the two populations are not thought to be the same e.g., spectral indices derived from the two emission mechanisms tend to differ \citep{silva2000, white2011}. However, in our case the similar indices and the spatial/temporal relationship between the X-ray and radio emission provide an opportunity of estimating the number density of non-thermal electrons ($N$) from X-ray and exploring the use of this value in Equation~\ref{E-nu_peak}.

\subsubsection{Estimating X-ray emission volumes, electron densities, and magnetic field strength}
	\label{S-Xrays}

As mentioned, the goal is to find an estimate of the non-thermal density of electrons $N$ from the X-ray analysis, and use this value in Equation~\ref{E-nu_peak} to estimate the magnetic field strength (we reiterate that this is based on the assumption that $N$ from an X-ray analysis can be used in the \citet{dulk1982} formulation, which we discuss in Section~\ref{S-number_dsicussion}). The electron number density for both the thermal and non-thermal thin-target components of the X-ray emission can be estimated, respectively, from
\begin{equation}
n_{0}=\sqrt{\frac{\xi}{V_0}} \\\\\\\\\\\\\\\\\\\\\\\\\\\\\\\\ \big[\mathrm{cm}^{-3}\big]
\label{eqn:em_measure}
\end{equation}
\begin{equation}
N=\frac{[n_{0}V_0\overline{F}]}{n_{0}V_{nth}}\frac{\delta_{xray}-1}{\delta_{xray}-0.5}E_{min}^{-1/2}\sqrt{ \frac{m}{ 2 } } \bigg( \frac{E_0}{E_{min}} \bigg)^{\delta_{xray}-1} \\\\ \big[\mathrm{cm}^{-3}\big]
\label{eqn:N_eq}
\end{equation}
\citep{mussetYYYY} where $\xi$ is the emission measure, $V_0$ and $V_{nth}$ are thermal and non-thermal X-ray source volumes, respectively, $N$ is non-thermal electron number density, $[n_{0}V_0\overline{F}]$ and $\delta_{xray}$ are as above, $E_{0}$ is the energy above which there is a power-law distribution, $E_{min}$ is the energy above which the electron density is to be calculated, and $m$ is electron rest mass expressed in keV/c$^2$. $[n_{0}V_0\overline{F}]$, $\delta_{xray}$, and $E_{0}$ are known from the above thermal and thin-target fit to the X-ray flux spectrum. We set $E_{0}=E_{min}$ so as to calculate the electron number density in the entire power law distribution. 

{The remaining unknown values are the volume $V_{nth}$ and $V_{0}$ of the thin-target and thermal sources, respectively. Fortunately, GOES SXI provides us with the opportunity to make an initial estimate of the thermal X-ray emission volume. The SXR extended source radius is $\sim$0.25\,R$_{\odot}$; assuming a simple spherical volume for this source we find $V_0$$\sim$$2.2\times10^{31}$\,cm$^{3}$. This is quite large, and we choose this to represent an upper limit on the source size. As for the non-thermal source, we choose standard HXR source volumes in previous results. For example, non-thermal X-ray sources from electron acceleration regions at coronal looptops have been studied by \citet{xu2008}, showing source size can be up to $15\,''$, ($5\times10^{27}$\,cm$^3$, assuming a spherical acceleration region). \citet{jeffrey2014} showed that the observed length (or volume) of a non-thermal source can be modelled assuming an electron acceleration region width of $23\,''$ (on the order of $10^{27}$\,cm$^3$, assuming a spherical geometry). \citet{krucker2010} and \citet{krucker2014} observed an above the looptop HXR source from an electron acceleration region and found the source volume to be $0.8\times10^{27}$\,cm$^3$ and $0.7\times10^{27}$\,cm$^3$, assuming a cylindrical and spherical volume, respectively. X-ray imaging observations have also observed thermal X-ray sources in solar flares to be up to $10^{27}-10^{28}$\,cm$^{-3}$ \citep{simoes2013, warmuth2013a, warmuth2013b}. Because we do not know the exact volumes of the thermal and non-thermal sources, we explore a range of physically reasonable values of $V_{0}$=$V_{nth}$=$10^{27}-10^{32}$\,cm$^{3}$, motivated by a lower limit based on previous estimates of HXR volumes and an upper limit based on what we observe in GOES SXI. The range in volumes results in a range for the ambient thermal and non-thermal electron densities ($n_0$ and $N$ from Equations~\ref{eqn:em_measure} and \ref{eqn:N_eq}). This range for $N$ is then used in Equation~\ref{E-nu_peak} to calculate a range of magnetic field strengths.}


\begin{figure}[!t]
    \begin{center}
    \includegraphics[scale=0.34, trim=1.2cm 0cm 0cm 0cm]{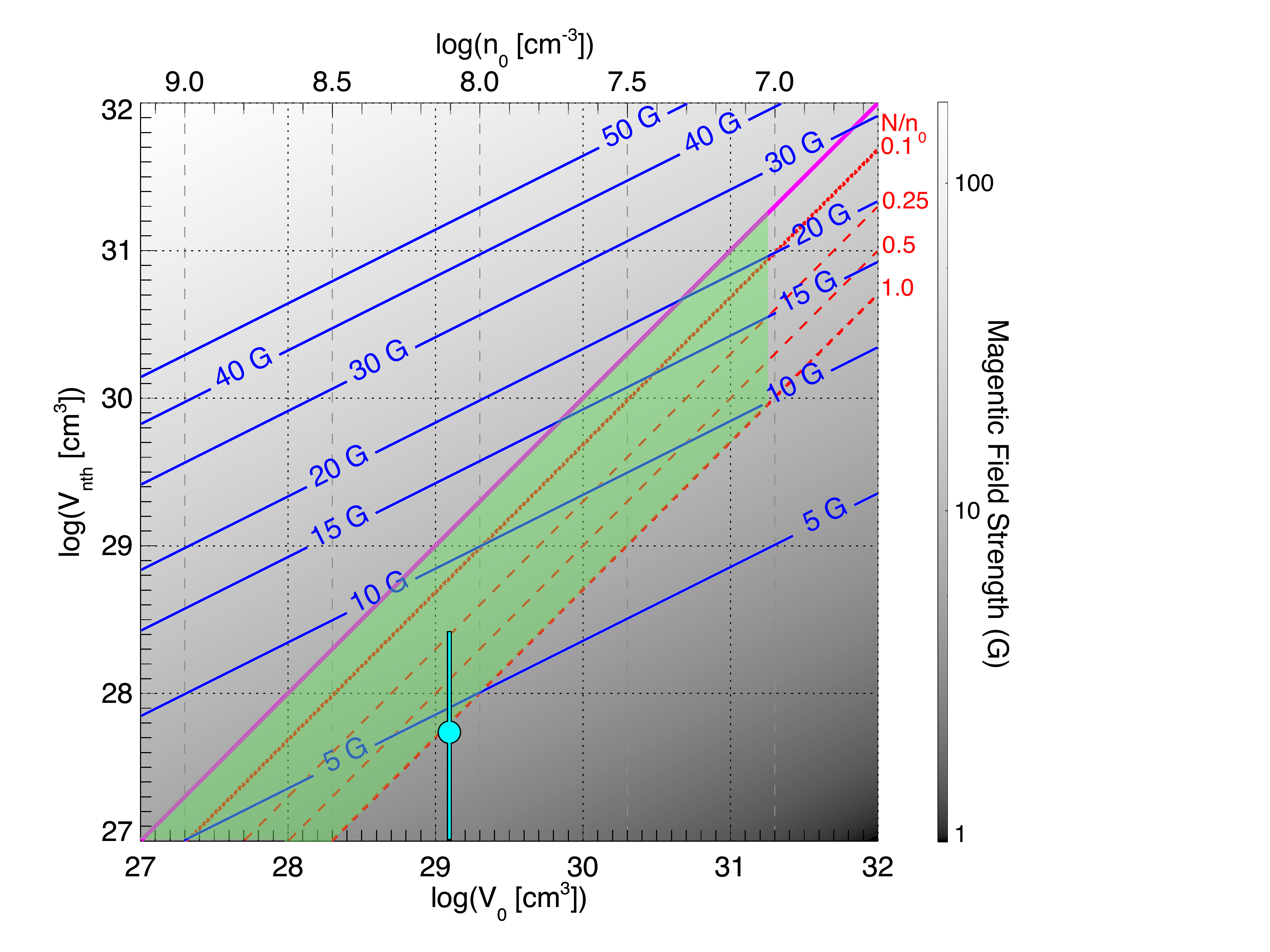}
    \caption{{Parameter space of magnetic field strength as a function of thermal and non-thermal X-ray volumes, with the blue contours indicating specific magnetic field strength. The pink line indicates equal thermal and non-thermal source volumes. The top x-axis indicates $n_0$ values, which depend directly on $V_0$ (via Equation~\ref{eqn:em_measure}). The red lines indicate the volumes at which specific ratios of thermal to non-thermal electron densities ($N/n_0$) occur (as indicated). The green shaded area marks the range of magnetic fields possible, given particular constraints on the volumes and ratios of $N/n_0$. The blue circle with the error bar is the magnetic field strength calculated from the numerical model below.}}
    \label{fig:Bfield-space}
    \end{center}
\end{figure}

Since the calculated magnetic field strength ultimately depends on \emph{both} the thermal and non-thermal volumes, we are left with a 2-dimensional space of magnetic field solutions, as plotted in Figure~\ref{fig:Bfield-space}. The shaded grey surface is the magnetic field strength (with blue contours indicating specific values) as a function of thermal and non-thermal volumes. The diagonal pink line indicates equal thermal and non-thermal volumes. The upper x-axis indicates the thermal electron densities calculated from Equation~\ref{eqn:em_measure}, and the red lines indicate volumes at which the ratio of non-thermal to ambient thermal electron density has values of $N/n_0=0.1,~0.25,~0.5,~1.0$, as indicated. This allows us to explore a reasonable estimate for the magnetic field strength as a function of thermal and non-thermal volume, subject to constraints on relative volume sizes, and realistic ratios of $N/n_0$. For example, we would not expect the non-thermal volume to exceed the thermal volume by a great extent, especially when both ambient thermal and non-thermal electrons are within a singular eruptive structure. Hence values of magnetic field above the pink line may be excluded as unreasonable. As well as this, previous studies have indicated ratios of $N/n_0<0.1$ \citep{gary1985, bain2014}. In extreme cases values as high as $N/n_0=1.0$ have been found \citep{krucker2014}, but values in excess of this are likely to be physically unreasonable; hence magnetic field strengths below the $N/n_0=1.0$ red line are also ruled out. Hence, the values of reasonable volumes and ratios of $N/n_0$ define a range in this space of reasonable magnetic field strengths. Given $V_{0}$=$V_{nth}$=$1\times10^{27}-2.2\times10^{31}$\,cm$^{3}$, $V_{nth} \leq V_0$ and a typical value of $N/n_0\leq1$, we find a range of magnetic field estimates of $\sim$$B$=4--25\,G, {as indicated by the shaded green region in Figure~\ref{fig:Bfield-space}}. While this range is quite large, it nonetheless falls close to the ranges of 1--15\,G from previous estimates in CMEs $\sim$1.5\,R$_{\odot} $\citep{bain2014, tun2013}. 

{In summary of Section~\ref{S:DM_approx}, the goal was to estimate a value for magnetic field strength from Equation~\ref{E-nu_peak} using estimates of $\nu_{peak}$, $\delta_{radio}$, $\theta$, $L$ and $N$ as defined above. These parameters were estimated as follows:}

\begin{enumerate}
\item Values for $\delta_{radio}\sim3.2$ and $\nu_{peak}\sim1$\,GHz were estimated from a fit of Equation~\ref{eqn:flux_density_fit} and ~\ref{eqn:alpha_thin} to the radio flux density spectrum constructed from NRH and RSTN.

\item A value for $L\sim0.45$\,R$_{\odot}$ was estimated from average source size in radio images $>$327\,MHz and a value of $\theta\sim45^{\circ}$ was assumed given the low polarisation of the type IV (and given that this would be the average angle along the LOS when looking through a complicated magnetic structure; {\color{black}generally, changing $\theta$ by a large amount did not affect the results significantly}).

\item To calculate the number density of non-thermal electrons $N$ (the remaining unknown in Equation~\ref{E-nu_peak}), Equation~\ref{eqn:N_eq} was used. The inputs into this equation were from a fit of a thin target model to the FERMI GBM spectrum i.e., $\xi\sim2.0\times10^{45}$\,cm$^{-3}$, $E_0\sim E_{min}\sim$9\,keV, $\delta_{xray}\sim2.9$,   and $n_{0}V_0\overline{F}\sim7.2\times10^{53}$\,electrons\,cm$^{-2}$\,s$^{-1}$\,keV$^{-1}$ . The input thermal and non-thermal volumes were given a range of $V_{0}$=$V_{nth}$=$1\times10^{27}-2.2\times10^{31}$\,cm$^{3}$; this gave a range for $N$.

\item Using the range for $N$ and values $\delta_{radio}$, $\nu_{peak}$, $L$, and $\theta$, a range of $B$ was then calculated from Equation~\ref{E-nu_peak}. Because $N$ depends on ranges of $V_0$ and $V_{nth}$ the possible values for $B$ is represented as a 2D space in Figure~\ref{fig:Bfield-space}. From reasonable assumptions that $V_{nth}\leq V_0$ and $N/n_0\leq1$, $B$ is then restricted to $B\sim4-25$\,G, as indicated by the green shaded region. We show in the Appendix why the contours of constant $B$ have a slope of 0.5 in the log-log space of $V_{nth}$ versus $V_{0}$.

\end{enumerate}
Ultimately, this result depends on the estimate of the non-thermal electron density $N$ derived from X-ray observations, which in turn depends on the estimate of X-ray source volumes; this assumes that both radio and X-ray emissions are from the same population of electrons within the CME, and that the CME plasma environment is homogenous in each property. {\color{black}The analysis provides the rare possibility of comparing X-ray and radio emitting CME electrons, which we discuss further in Section~\ref{S-Conclusion}.}
We next test the validity of the B-field values calculated in this way by comparing such results to a numerical model for gyrosynchrotron radiation.

%
%
\subsection{Magnetic field and non-thermal electron property estimates from a numerical model of gyrosynchrotron radiation}

In the previous section, we used a parametric fit to the radio flux density spectrum and the \citet{dulk1982} approximations to estimate the magnetic field strength of the CME. In order to check the validity of these results, and also estimate a more accurate value for the magnetic field, we next try numerical modelling of the gyrosynchrotron flux density spectrum.

We employ full gyrosynchrotron numerical calculations based on the formalism originally outlined in \citet{ramaty1969}. We use the numerical code developed by \citet{simoes2006} and improved for speed by \citet{CostaSimoesPinto:2013}. The code solves for the gyrosynchrotron emission $j_{\nu}$ and self-absorption $\kappa_{\nu}$ coefficients, and radiative transfer for the ordinary and extraordinary magneto-ionic modes and includes the effects of Razin suppression, considering a uniform cylindrical flare source. The model takes a variety of parameters as input including the characterisation of the source and the distribution of the non-thermal electrons. The source is characterised by its magnetic field strength $B$, source angular diameter size $\Lambda$ and length along the line-of-sight (LOS) $L$, background density of thermal electrons $n_0$, angle between the line-of-sight and the magnetic field $\theta$; the non-thermal electron populations is defined as a power-law distribution with a number density $N$, energy spectral index $\delta$, and electron energy range $[E_0, E_1]$. In using this model we assume that all parameters are homogenous throughout the medium.

\begin{figure}[!t]
    \begin{center}
    \includegraphics[scale=0.4, trim=1cm 5cm 0cm 5cm]{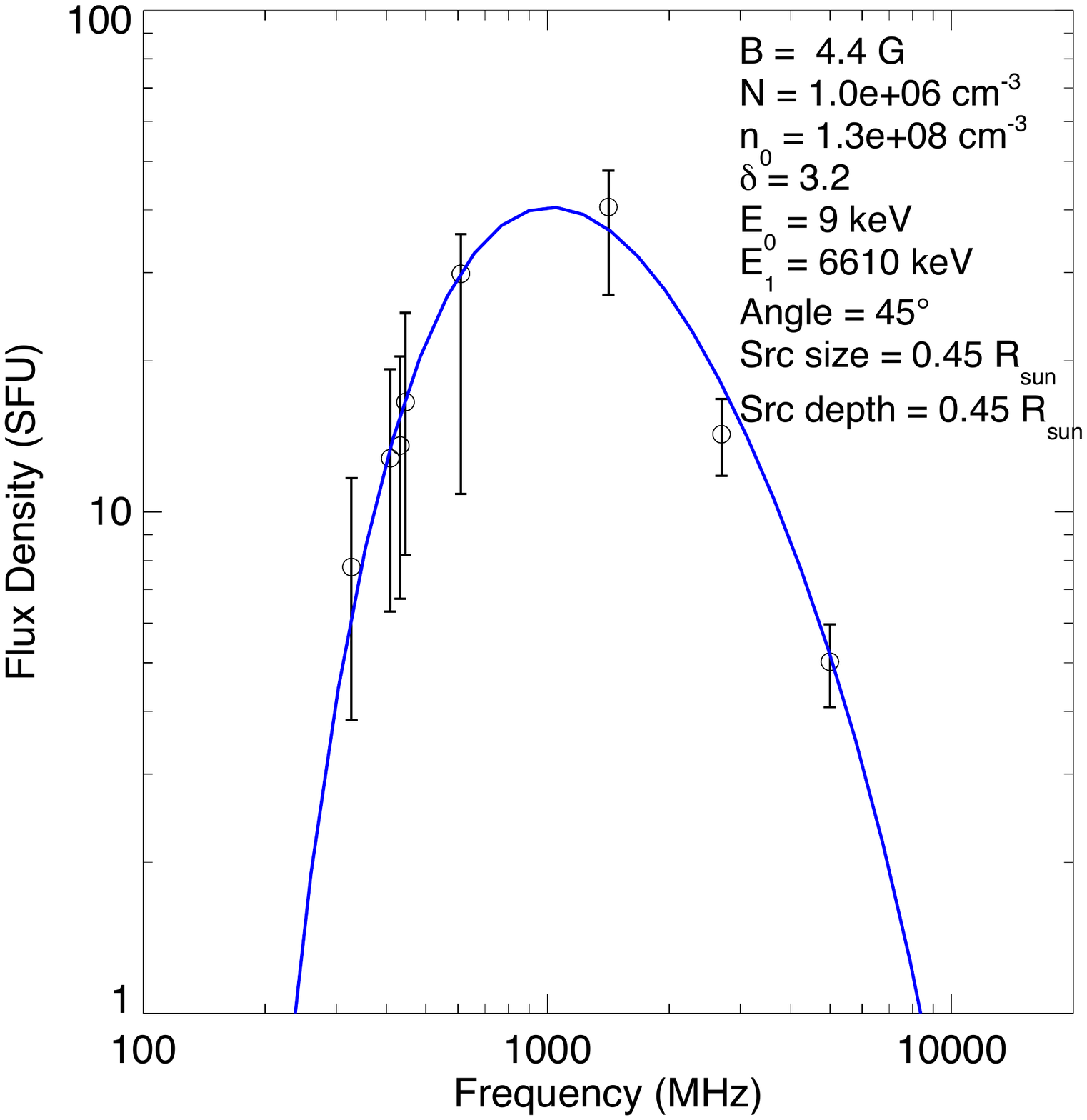}
    \caption{{Gyrosynchrotron model fit to the average flux density data from RSTN and NRH between 11:01:10--11:03:40\,UT. The model parameters are indicated in the legend.}}
    \label{fig:gyro_model}
    \end{center}
\end{figure}

{Since the model has a large number of parameters, and we have a small number of data points, we choose to fix some of the parameters to avoid over-fitting the spectrum. We fix the source size and LOS length fixed at $\Lambda=L=0.45$\,R$_{\odot}$ i.e., the average width of the semi-major axis of the sources between 327--445\,MHz throughout their lifetime. As before, the angle between our LOS and B-field is unknown so we fix it an average of 45$^{\circ}$. From Section~\ref{S:DM_approx}, we set the lower cut-off energy of $E_0$=9\,keV and choose a $\delta=3.2$. The remaining fit parameters are $B$, $N$, $n_0$, $E_1$. We use the Levenberg-Marquardt least-squares method to fit the parameterised gyrosynchrotron model to the average flux density spectrum between 11:01:10--11:03:40\,UT. We use starting parameters of $B=5$\,G, $n_0\sim10^8$\,cm$^{-3}$ (this is assuming the plasma frequency source in Figure~\ref{fig:aia_nrh}(a) comes from a similar location to the gyro-emission source), $N(>E_0)=0.1n_0=1\times10^7$\,cm$^{-3}$ and $E_1=7$\,MeV. The fit fit spectrum is plotted in Figure~\ref{fig:gyro_model}, with the fit values listed in Table~\ref{table:results} (the uncertainties are the 1-sigma errors from the fit). The resulting magnetic field strength is $4.4\pm2.7$\,G. } 

\begin{table}[!t]
\begin{center}
\begin{tabular}{l*{6}{c}r}
\hline
	$B_f$ & $N_f$ & $n_{0,f}$ & $E_{1,f}$ \\
	(G)     & (cm$^{-3}$) & (cm$^{-3}$) &  (MeV) \\
\hline
\hline
	 $4.4\pm2.7$   &  $(1.0\pm0.7)\times10^6$ &  $(1.3\pm0.2)\times10^8$ & $6.6\pm0.2$  \\
\hline
\end{tabular}
\caption{{Resulting parameters and uncertainties from a fit of the numerical model to the flux density spectrum.}}
 \label{table:results}
\end{center}
\end{table}

{Interestingly we can make a direct comparison between the numerical model and the above analysis from the \citet{dulk1982} approximations. The position of $\sim$4.4$\pm$2.7\,G at $n_0=1.3\times10^8$\,cm$^{-3}$ is marked on Figure~\ref{fig:Bfield-space} and falls close the magnetic field range as calculated by the \citet{dulk1982} approximations above (the green shaded region in the figure).
Despite the first method giving only a range of magnetic field estimates, there is some level of consistency between the results. However, the numerical model requires a value for $N$ an order of magnitude lower than that used in the \citet{dulk1982} formulation, we discuss this further in Section~\ref{S-number_dsicussion}}

\section{Discussion}
     \label{S-Conclusion}                

The observations of simultaneous radio and X-ray spectroscopic/imaging observations of a behind the limb event are relatively rare, providing an opportunity to diagnose the magnetic field strength in a radio source which we assume to be within the CME (see below). This was done using both the \citet{dulk1982} approximations and a numerical model. While the \citet{dulk1982} formulation combined with analysis of FERMI GBM gave a range of possible values of $B=4-25$\,G, use of the more accurate numerical model gave values of $4.4\pm2.7$\,G. 
{\color{black}These methods also provide a rare opportunity of comparing the properties of the radio and X-ray emitting CME electrons, which we discuss here.}

\subsection{Comparison of electron distribution properties from X-ray and radio observations}
	\label{S-number_dsicussion}

{In general, radio and X-ray emissions are thought to be from separate populations of energetic electrons. This is mainly due to the fact that electron spectral indices derived from X-ray and radio observations may differ \citep{silva2000}. This is interpreted as the two types of emission belonging to spatially separated electron populations \citep{gary1994} and/or electron populations at different energy ranges \citep{white2011, marsch1994, trottet1998, trottet2015}, with the radio emission generally coming from higher energy electrons than the X-ray. 
However, several authors have found similarities in the X-ray and microwave emitting electrons properties of solar flares \citep{nitta1986, klein1986, wang1994}. In a study similar to ours, \citet{krucker2010} presented a behind the limb event associated with a coronal thin-target HXR source at 16--80\,keV and gyrosynchrotron spectrum with low turnover frequency of 2 GHz. They found both types of emission could be produced from a single power-law population of electrons of spectral index 3.4 in coronal flare loops.

In this study we find that the electron spectral indices derived from radio and X-ray are similar, with values of $\delta_{radio}$=$3.2\pm0.3$ and $\delta_{xray}=2.9\pm0.5$, see Figure~\ref{fig:spectral_indices} (note they represent a negative slope). This provided evidence that both emissions were produced by similar electron populations. However, as mentioned, it is likely that the two emission types come from different electron energy ranges. Indeed we can explicitly test such an assertion here.

The numerical model allows us to explore the possible energy ranges of the non-thermal electron distribution which produce a reasonable model fit to our radio data. The gyrosynchrotron model fit to our data shows the electron energy range to be from $E_0=9$\,keV to $E_1=6.6$\,MeV, however the observed spectrum can also be fit with a minimum energy cutoff as high as $E_0=1$ MeV, indicating that it may only be electrons above this energy range which contribute to the radio emission.
For $E_0>1$\,MeV the flux in the optically thick part of the spectrum starts to decrease and fails to match the observed values; for $E_1$$<$6.6\,MeV again the calculated flux in the optically thin part of the spectrum begins to fall below the observed values. Moreover, the Razin suppression \citep{ramaty1969} is quite strong, as indicated by the Razin parameter $\alpha_R=1.5\nu_B/\nu_p=0.21$ using the values for $B$ and $n_\mathrm{0}$ from the model. The gyrosynchrotron emission from an electron with a Lorentz factor $\gamma$ is strongly suppressed at the lower frequencies if $\alpha_R \gamma < 1$. Thus, for $\alpha_R=0.21$, the lower frequency radiation from electrons with $\gamma \lesssim 4.8$ (i.e. $E < 1.9$ MeV) is strongly suppressed, indicating that non-relativistic electrons may not be important in producing the gyrosynchrotron emission in this event. {\color{black}These energies are an order of magnitude larger than the electrons that produce the X-rays (on the order of $\sim$10\,keV).}

We can also make a comparison of the number densities of radio and X-ray electrons. Using $E_0=1$\,MeV as the minimum energy of the non-thermal electrons that produce the radio emission, the model gives a non-thermal number density of $N(\textgreater1\,\mathrm{MeV})\approx50$\,cm$^{-3}$. This would suggest that the number density of radio emitting electrons is far lower than those emitting X-ray e.g., we calculated $N(\textgreater1\,\mathrm{MeV})\approx50$\,cm$^{-3}$ for radio emitting electrons, and in Section~\ref{S-Xrays} a reasonable ratio of $N/n_0\sim0.1$ would give a range of $N(\textgreater9\,\mathrm{keV})\approx10^6-10^8$\,cm$^{-3}$ for the X-ray emitting electrons. So how do we justify the use of the values of $N(\textgreater9\,\mathrm{keV})$ from X-ray analysis in the \citet{dulk1982} radio formulation? This is because this radio formulation accounts for electrons with energies as low as $\sim$$10$\,keV, a refractive index of unity and no Razin suppression i.e. it takes into account radio radiation from all energetic electrons. While the large $N$$\sim$10$^7$\,cm$^{-3}$ (which includes electrons as low as 9\,keV) may also be input into the numerical model, the results of the model show that such electrons contribute a negligible amount of radiation when a full treatment of Razin suppression is accounted for. If the \citet{dulk1982} formulation accounted for such effects, $N(\textgreater9\,\mathrm{keV})$ calculated from X-rays could not be used in such a radio formulation (Equation~\ref{E-nu_peak}).

In total, the above analysis suggests that the radio and X-ray emission in this event come from a similar population of electrons of spectral index $\delta\sim3$, with the X-rays originating in the lower energy part of the distribution (on the order of $\sim$10\,keV), and the radio originating in the higher energy part of the distribution (on the order of $\sim$\,1\,MeV).

\subsection{Location and lifetime of energetic electrons in the CME}
\label{S-lifetime-location}

{Figure~\ref{fig:cme_radio_expansion} provides good evidence that the expansion of the radio source closely follows the expansion of the internal structure of the CME, especially toward the south. This suggests that the radio sources (and energetic electrons) are located on the internal magnetic field of the eruption.
Figure~\ref{fig:fermi_gbm}(a)-(c) also shows that the radio emitting region is always contained in the SXR emitting region as it grows, which likely demarcates at least part of the erupting structure. Therefore, given the similar kinematics and morphology seen between radio and both the EUV and SXR signatures of the CME, we reason that the energetic electrons are located on the internal magnetic field of the eruption.} 

As for the lifetime of the energised electrons within the CME, knowledge of energy ranges for electrons producing radio emission also allows us to analyse the electron thermalisation time and determine if the radio emission diminishes due to thermalisation alone. If the energetic electrons are trapped in the magnetic field of the CME, their collisional lifetime can be estimated by assuming that they will thermalise after crossing a column depth of $N_\mathrm{stop}=10^{17}E^2$\,cm$^{-2}$, where $E$ is expressed in keV \citep{Tandberg-HanssenEmslie:2009}. Making $N_\mathrm{stop}=n_\mathrm{0}L$, this results in the thermalisation time of $t=L/v=10^{17}E^2/n_{0}v$. With $n_\mathrm{0}=1.3\times 10^8$ cm$^{-3}$ from the gyrosynchrotron analysis and $E=1000$\,keV ($v\approx 0.94c$), we find $t\approx7.6$\,hours. The energy dependent Coulomb collisional loss rate from \citet[][; their Equation 4]{vilmer1982} gives a thermalisation time of 5.6 hours for 1\,MeV electrons in a background plasma of electron density of $n_\mathrm{0}=1.3\times 10^8$ cm$^{-3}$.
Both methods give consistent results, and such times are of course much longer than the duration of the type IV radio burst in Orf\'{e}es ($\sim$5 minutes), suggesting that the radio emission decreases due to (i) the electrons escaping the radio emitting environment to regions of lower magnetic field strength, meaning they no longer emit gyrosynchrotron emission efficiently\footnote{This idea would be consistent with energetic particles precipitating to the solar surface, as outline in \citet{ackermann2017}}; or (ii) the electrons remain trapped in the expanding CME, and thus the number density of the non-thermal electrons reduces as the CME expands, making the gyrosynchrotron emission less efficient. Given that the radio sources expands simultaneously with the CME, as shown in Figure~\ref{fig:cme_radio_expansion}, effect (ii) is a possibility. 

{Finally, during the lifetime of the radio burst there is no frequency change of the type IV radio burst. In general, observations of type IVs do not necessarily show a frequency drift \citep{pick1986}. This is despite the fact that Equation 2 would suggest that changing $N$ and $B$ in the CME environment would result in changing $\nu_{peak}$. However, given the small amount of free magnetic energy (5--20\% of total magnetic energy budget) used in driving the initial CME launch \citep{forbes2000, devore2005, roussev2012}, the total magnetic field strength (free energy plus potential) may not decrease by a large amount during launch, resulting in only small changes in $\nu_{peak}$ due to magnetic energy expenditure (especially during the 5 minute window of our observation). Furthermore, $\nu_{peak}$ is relatively insensitive to large changes in $N$ and also has a dependency on $L$ and $\theta$, which can change in various ways and lead to an increasing or decreasing $\nu_{peak}$. Therefore, constant $\nu_{peak}$ does not necessarily imply constant conditions in the CME plasma.}

\subsection{CME magnetic and mechanical energies}   

If we assume that the magnetic field strength of the CME is 4.4\,G throughout its volume, then we may estimate the total magnetic energy content from $E_{mag}=V_{cme}B^2/8\pi$. With a spherical volume given by $V_{cme}=4/3\pi L_{cme}^3$, where $L_{cme}=$0.6\,R$_{\odot}$ e.g., the approximate diameter of the CME seen in AIA images at the time of observation of the radio sources (this size is larger than the gyro-emission radio sources).
We then find the magnetic energy to be $E_{mag}=6.4\times$10$^{32}$\,ergs. While the assumption of homogenous magnetic field in the CME is quite a simplified one, the value is similar to the magnetic energies (calculated from non-potential field extrapolations) reported in \citet{emslie2012}. It is also a more direct estimate of the magnetic energy than that provided in \citet{vourlidas2000}, which used in-situ measurements for magnetic clouds to estimate the field strength low in the corona.

This magnetic energy may then be compared to the mechanical energy (kinetic and potential). From \citet{pesce-rollins2015}, the CME mass is $M_{cme}=1\times10^{16}$\,g, and with a velocity of 2000\,km\,s$^{-1}$, giving a kinetic energy is $2.0\times10^{32}$\,ergs. Similarly, the CME potential energy at a distance approaching infinity is $2.0\times10^{31}$\,ergs, giving a total mechanical energy of $2.2\times10^{32}$\,ergs. {The calculation of total mechanical energy firstly assumes the CME has a constant mass from the time of launch at the solar surface (as in \citet{aschwanden2009}), and also propagates radially from the solar surface (1\,R$_{\odot}$) to infinity, as in \citet{vourlidas2000}}. This CME was particularly massive and fast, so the mechanical energy represents 34\% of the total magnetic energy of the CME in the corona i.e., a large amount of the CME magnetic energy is expended in lifting the CME from the gravitational potential of the sun and accelerating it to 2000\,km\,s$^{-1}$. We emphasise here that this calculation includes both the non-potential and potential magnetic energy content of the CME, since we have no information of the proportionality between the two e.g., the non-potential energy may represent just a small amount of the total magnetic energy (between 5-20\% \citep{forbes2000, devore2005, roussev2012}), but of course must be at least as large as total mechanical energy. The magnetic energy dominating the total energy content of the CME has been found in previous observations \citep{vourlidas2000, emslie2012, carley2012}.

\section{Conclusion}

This study detailed the observation of a large behind the limb flare and CME associated with an extended radio type IV source observed by NRH and RSTN, being identified as a mixture of both plasma and gyrosynchrotron emission below 300\,MHz and gyrosynchrotron radiation above this frequency. The event was also associated with HXR observations from FERMI GBM and X-ray imaging observations of the eruption using GOES SXI. This unique set of X-ray and radio observations allowed us to diagnose both the CME magnetic field strength and a variety of energetic electron properties including number density, spectral index and energy ranges which contribute separately to the radio and X-ray emission. The magnetic field strength and various other properties were diagnosed through two main methods: 

\begin{enumerate}
\item Parametric fitting and the \citet{dulk1982} approximations: Using a parametric fit to the radio flux density spectrum and analysis of X-ray emissions from \emph{Fermi} GBM we showed that both the radio and X-ray emission came from a similar population of non-thermal electrons with a spectral index of $\delta$$\sim$3. The X-ray analysis was used to estimate a range for the number density of non-thermal electrons $N$, and this was used in the \citet{dulk1982} approximations to calculate a range of possible magnetic field strengths of 4-25\,G. This large range was ultimately due to the unknown source size and volume (and hence unknown $N$) of the X-ray emission, highlighting the importance and future necessity of X-ray imaging observations in coronal plasma diagnostics.

\item \citet{simoes2006} numerical model: In this second method we use a full numerical model for gyrosynchrotron radiation to fit our flux density spectrum and estimate a magnetic field of $\sim$4.4\,G, a much more accurate measurement of the field compared to the \citet{dulk1982} approximations. This method also allowed an estimate of the electron energy ranges involved in the gyrosynchrotron emission, placing them in the range of $\sim$1\,MeV to 6.6\,MeV. 
\end{enumerate}
A magnetic field strength of  $\sim$4.4\,G at a CME core height of $\sim$1.3\,R$_{\odot}$ is similar to the values of CME magnetic field previously found at a CME core e.g., \citet{bain2014}. Such diagnostics are an important part of CME dynamics and what ultimately drives the eruption.

In general, in both methods the NRH and RSTN observation were essential to this analysis, with RSTN providing a critical measurement of the non-thermal electron spectral index and NRH providing a measure of source size (and of course position relative to the CME). Further instrumentation should provide sensitive and calibrated flux density and imaging observations in a continuous frequency range from decameter to millimeter wavelengths in order to provide the possibility of further observing CME gyrosynchrotron spectra with improved accuracy. Such future instrumentation should also be capable of a large dynamic range, given that gyrosynchrotron flux densities can be orders of magnitudes below plasma emission flux densities.

Overall, gyrosynchrotron emission from radio bursts associated with CMEs is still relatively rare, possibly due to a lack of sensitivity and dynamic range. A frequent and regular imaging of gyro-emission from CMEs in the future will be invaluable in investigating the magnetic field of these eruptive events on a routine basis. Further effort should also be made in investigating the relationship between radio and X-ray emitting electron populations in flare and/or CME events. This may be possible in the observing synergies between future X-ray imagers such as the Spectrometer Telescope for Imaging X-rays (STIX) on board \emph{Solar Orbiter} and future radio interferometers such as the Square Kilometre Array (SKA) and the Low Frequency Array \citep[LOFAR;][]{vanHaarlem2013}.

Finally, while all studies to date give some suggestion as to the location of the magnetic field strength measurement within the CME e.g., front, core or flank, no information exists to date on the spatial distribution and relative strengths of the magnetic fields in different parts of a CME. Future observations or instrumentation should aspire to such measurements. For the moment, magnetic field measurements in CMEs remain the most important yet elusive diagnostic in CME physics.

\section*{Acknowledgments}
\noindent
EPC is supported by ELEVATE: Irish Research Council International Career Development Fellowship -- co-funded by Marie Curie Actions. PJAS acknowledges support from STFC Consolidated Grant ST/L000741/1. PJAS and NV would like to thank ISSI for the support of the team `Energetic Ions: The Elusive Component of Solar Flares'.
We are grateful to the SDO, Fermi, RSTN and GOES teams for open access to their data. The NRH is funded by the French Ministry of Education and the R\'{e}gion Centre. Orf\'{e}es is part of the FEDOME project, partly funded by the French Ministry of Defense. The authors acknowledge the Nan\c{c}ay Radio Observatory / Unit\'{e} Scientifique de Nan\c{c}ay of the Observatoire de Paris (USR 704-CNRS, supported by Universit\'{e} d'Orl\'{e}ans, OSUC, and R\'{e}gion Centre in France) for providing access to NDA observations accessible online at http://www.obsnancay.fr. N.V. would like to acknowledge the support of CNES and of the French Program on Solar-Terrestrial Physics of INSU (PNST). We would like to thank Sam Krucker for some useful discussion on the interpretation of the X-ray data. We would also like to thank the referee for their useful comments and suggestions on improving our analysis and the paper.

\section*{Appendix}
We wish to show that contours of constant $B$ in the log-log space of $V_{nth}$ versus $V_0$ have a slope of 0.5. Firstly, the dependency of $B$ on the volumes is via the dependency of $N$. From Equation 4 we have $n_0 \propto V_0^{-1/2}$ , and Equation 5 gives  

\begin{equation}
N \propto \frac{1}{n_0 V_{nth}}=\frac{V_0^{1/2}}{V_{nth}}
\end{equation}
Note that in Equation 5, the property $[n_{0}V_0\overline{F}]$ is a constant so $V_0$ in this expression does not enter the proportionality. Now, from Equation 2 we have $B^{0.68-0.03\delta} \propto N^{-0.32+0.03\delta}$. Using a value of $\delta=3$ and expression for $N$ above we then obtain
\begin{equation}
B\propto \Bigg(\frac{V_{nth}}{V_0^{1/2}}\Bigg)^{0.4} = \frac{V_{nth}^{0.4}}{V_0^{0.2}}
\end{equation}
Taking log$_{10}$ of all sides we obtain
\begin{equation}
\mathrm{log}_{10}(B) =  0.4\mathrm{log}_{10}(V_{nth}) - 0.2\mathrm{log}_{10}(V_0) 
\end{equation}
This means along lines of constant $B$ we have 
\begin{equation}
\mathrm{log}_{10}(V_{nth}) = 0.5\mathrm{log}_{10}(V_0) + C
\end{equation}
where C is a arbitrary constant. Hence in the log-log space of Figure 9, lines of constant $B$ have a slope of approximately 0.5.

\bibliographystyle{aa.bst}

\end{document}